\documentclass[sigplan,10pt,nonacm]{acmart}
\usepackage{ebproof}
\usepackage{mathtools}
\usepackage{ifthen}
\usepackage{natbib1}

\usepackage[capitalize,nameinlink]{cleveref}
\usepackage{tikz}
\usepackage{tikzpagenodes}
\usetikzlibrary{matrix,fit,quantikz,tikzmark}
\usepackage{pgfplots}
\usepgfplotslibrary{groupplots,colorbrewer}
\usepackage[inline]{enumitem}
\usepackage{moreenum}
\usepackage[T1]{fontenc}
\usepackage[utf8]{inputenc}
\usepackage{textcomp}
\usepackage{listings}
\usepackage{color}
\usepackage{xparse}
\definecolor{keywordcolor}{rgb}{0.7, 0.1, 0.1}   % red
\definecolor{tacticcolor}{rgb}{0.0, 0.1, 0.6}    % blue
\definecolor{commentcolor}{rgb}{0.4, 0.4, 0.4}   % grey
\definecolor{symbolcolor}{rgb}{0.0, 0.1, 0.6}    % blue
\definecolor{sortcolor}{rgb}{0.1, 0.5, 0.1}      % green
\definecolor{attributecolor}{rgb}{0.7, 0.1, 0.1} % red

\makeatletter
\def\addToLiterate#1{\edef\lst@literate{\unexpanded\expandafter{\lst@literate}\unexpanded{#1}}}
\lst@Key{add to literate}{}{\addToLiterate{#1}}
\makeatother
\makeatletter
\patchcmd{\@addmarginpar}{\ifodd\c@page}{\ifodd\c@page\@tempcnta\m@ne}{}{}%
\makeatother
\reversemarginpar
\newif\ifcomments\commentsfalse
\ifcomments
    \advance\marginparwidth \marginparsep
    \advance\marginparwidth -4pt
    \advance\marginparwidth 210mm
    \advance\marginparwidth -6.75in
    \long\def\remark#1{% for notes in the margin -- from Norman Ramsey
        \let\SAVEDspacefactor\relax
        \ifhmode\edef\SAVEDspacefactor{\the\spacefactor}\fi
        \raisebox{-3.5pt}[0pt][0pt]{\makebox[0pt][c]{$\scriptstyle\diamond$}}%
        \marginpar{\raggedright\hbadness=10000
        \parindent=8pt \parskip=2pt
        \def\baselinestretch{0.8}\footnotesize%\tiny
        \itshape\noindent #1\par}%
        \ifx\SAVEDspacefactor\relax\else\spacefactor\SAVEDspacefactor\fi\relax}
\else
    \long\def\remark#1{\relax}%
\fi

\AtBeginDocument{%
  }

\setcopyright{acmlicensed}
\copyrightyear{2026}
\acmYear{2026}
\acmDOI{XXXXXXX.XXXXXXX}

\acmJournal{JACM}
\acmVolume{37}
\acmNumber{4}
\acmArticle{111}
\acmMonth{8}

\DeclareMathAlphabet{\mathbbold}{U}{bbold}{m}{n}
\DeclareFontFamily{OT1}{pzc}{}
\DeclareFontShape{OT1}{pzc}{m}{it}{<-> s * [1.10] pzcmi7t}{}
\DeclareMathAlphabet{\mathpzc}{OT1}{pzc}{m}{it}

\newcommand{\data}[3]{\mathit{#1}_{\mathit{#2}}^{\mathit{#3}}}
\newcommand{\wmid}{\;\mid\;}

\newcommand{\denot}[1]{\llbracket{#1}\rrbracket}

\newcommand{\dom}[1]{\textnormal{funs}(#1)}
\newcommand{\TyIsoIso}[3]{
  \ifthenelse{\equal{#1}{\varnothing}}{%
    \vdash_{\mathrm{i}} #2 : #3%
  }{%
    #1 \vdash_{\mathrm{i}} #2 : #3%
  }%
}

\newcommand{\TnIso}[1]{}%{I-{#1}}

\newcommand{\einsum}{\mathop{\mathsf{einsum}}}
\newcommand{\expand}{\mathop{\mathsf{expand}}}

\newcommand{\bind}[3]{\mathop{\mathsf{let}} {#1} = {#2} \mathop{\mathsf{in}} {#3}}
\newcommand{\bindop}{\mathop{\mathsf{let}}}
\newcommand{\lookup}{\mathop{\texttt{lookup}}}
\newcommand{\disjoint}{\mathrel{\bot}}

\NewDocumentCommand{\compu}{ O{\ensuremath{\Gamma}} m m }{%
  \textit{comp}_{#1}(#2, #3)
}
\NewDocumentCommand{\shapeu}{ O{\ensuremath{\Gamma}} m }{%
  \textit{shape}_{#1}(#2)
}
\NewDocumentCommand{\gendim}{ m }{%
  \textit{gendim}(#1)
}
\NewDocumentCommand{\createdict}{ m }{%
  \textit{create\_dict}(#1)
}
\NewDocumentCommand{\rename}{ m m }{%
  \textit{rename}(#1, #2)
}

\NewDocumentCommand{\ivar}{s m m}{%
  \IfBooleanTF{#1}{%
    {#2}%
  }{%
    {#2^{(#3)}}%
  }%
}
\newcommand{\iname}[2]{\ivar*{#1}{#2}}
\NewDocumentCommand{\ivw}{s}{%
  \IfBooleanTF{#1}{%
    \ivar{w}{3}%
  }{%
    \iname{w}{3}%
  }%
}
\NewDocumentCommand{\ivx}{s}{%
  \IfBooleanTF{#1}{%
    \ivar{x}{2}%
  }{%
    \iname{x}{2}%
  }%
}
\NewDocumentCommand{\ivy}{s}{%
  \IfBooleanTF{#1}{%
    \ivar{y}{2}%
  }{%
    \iname{y}{2}%
  }%
}
\NewDocumentCommand{\ivz}{s}{%
  \IfBooleanTF{#1}{%
    \ivar{z}{2}%
  }{%
    \iname{z}{2}%
  }%
}

\begin{document}

\title{Formalizing $\einsum$ with Named Tensor Notation}

\author{Yafei Yang}
\email{yafyang@iu.edu}
\orcid{0009-0006-1402-8014}
\author{Chung-chieh Shan}
\authornotemark[1]
\email{ccshan@iu.edu}
\affiliation{%
  \institution{Indiana University}
  \city{Bloomington}
  \state{Indiana}
  \country{USA}
}

\begin{abstract}
% what's the problem we are solving? einsum opt
Einstein notation ($\einsum$) is a widely adopted tensor operation that has proven useful in areas such as machine learning and quantum computing.
But it had no formal semantics until recently, and its properties had no mechanized verification.
We formalize the denotational semantics of a language of tensors (multi-dimensional arrays) that supports $\einsum$ and repeated tensors ($\expand$).
With this semantics, we verify equational rewriting rules about $\einsum$ and $\expand$ that improve the performance of tensor computation.
Whereas typical languages use unnamed tensors, whose shape is a sequence of natural numbers, we found it advantageous for mechanization to use named tensors, whose shape is a finite map from names to natural numbers.
To our knowledge, this is the first time that the semantics and rewriting rules of $\einsum$ are mechanized in a proof assistant.
Our solution is agnostic to proof assistants; we have implemented it in Lean and partially in Agda.
\end{abstract}





\maketitle

\section{Introduction}
\label{section:intro}

Originating from physics as a shorthand on paper, \emph{Einstein notation} has since been implemented in libraries for practical computation and found expanded use in popular domains such as
machine learning \cite{Ansel_PyTorch_2_Faster_2024,harris2020array,tensorflow2015-whitepaper}, probabilistic programming \cite{chiangFactorGraphGrammars2024,dusellBdusellSemiringeinsum2026}, and quantum computing \cite{grayQuimbPythonPackage2018,bayraktarCuQuantumSDKHighPerformance2023}.
Also known as $\einsum$, it~can be thought of as a generalized multiplication that produces a tensor (a multi-\hspace{0pt}dimensional array) from a list of tensors.
It~is versatile: for instance, it~subsumes elementwise, matrix, and vector inner and outer product.
Indeed, it~can operate on any number of tensors, each with any number of dimensions.
These dimensions are \emph{named} in order to specify which sum-product, indexing, or broadcasting operations to perform \cite{wenigSyntaxSemanticsEinsum2025}.
For example, given an input matrix~$M$, we can compute the matrix product~$MM^{T}$ by specifying $\einsum(M_{ij}, M_{kj} \to ik)$ using the names $i,j,k$.

Unfortunately, the many implementations of $\einsum$ do not agree entirely with each other.
For instance, only some implementations allow multiple dimensions of the same input tensor to receive the same name.
In~PyTorch \cite{Ansel_PyTorch_2_Faster_2024}, we can extract the diagonal of a square matrix~$N$ by writing $\einsum(N_{ii}\to i)$, but the same expression causes a runtime error in another library Semiring Einsum \cite{dusellBdusellSemiringeinsum2026}.
Similarly, only some implementations allow multiple output dimensions with the same name.
According to \citet{wenigSyntaxSemanticsEinsum2025}, we can make a diagonal matrix from a vector~$V$ by writing $\einsum(V_{i}\to ii)$, but this expression is rejected by many implementations.
These inconsistencies arise because library implementers follow an intuitive understanding of what $\einsum$ does, but there was no formal definition of what counts as a valid $\einsum$ expression.

To address these inconsistencies, researchers recently proposed a formal semantics for $\einsum$ \cite{wenigSyntaxSemanticsEinsum2025}.
They specify how to treat multiple dimensions with the same name.
Furthermore, they use their semantics to prove a couple of rewriting rules for $\einsum$, making it possible for future implementers to optimize $\einsum$ computations.
But their formalization is not mechanized.
Moreover, they do not discuss how $\einsum$ interacts with other common tensor operations.

In the present work, we mechanize the syntax and semantics of $\einsum$.
In doing so, we face challenges presented by the semantics of tensors and the syntax of names.
Traditionally, the dimensions of a tensor are \emph{ordered} and \emph{unnamed}.
Thus, the shape of a tensor and an index into a tensor are both lists of numbers.
To~operate on such a tensor, each $\einsum$ would introduce arbitrary names and associate them with the ordered dimensions.
It~is in each $\einsum$---not in each tensor---that names like $i,j,k$ above are associated with dimensions.
In~order to specify well-formed $\einsum$ expressions, our mechanization would need to track these associations.
And because the output dimensions are ordered, it~would need to express the permutation of names.

To avoid these complications, we use \emph{named tensors} \cite{chiang2023named} instead.
The dimensions of a named tensor are \emph{unordered}, and \emph{named} independently of any operation.
Thus, the shape of a tensor and an index into a tensor are both finite maps from names to numbers.
Thanks to this representation, dimension names become intrinsic to tensors, so the syntax of $\einsum$ doesn't need to associate names to dimensions separately.
And because dimensions are unordered, names do not need to be permuted: a~tensor of shape $\{i\mapsto 3,k\mapsto 4\}$ is as good as a tensor of shape $\{k\mapsto 4,i\mapsto 3\}$, because these two shapes are simply the same finite map.

% TODO: Do we need to justify why we ignore repeated dimension names?
Based on this idea, we design a core tensor language with $\einsum$.
To~show that named tensors are just as expressive as unnamed tensors, we~translate unnamed tensor expressions to named ones.
To~study how to optimize $\einsum$ when it interacts with other tensor operations, we include another widely implemented operation $\expand$, which repeats the contents of a tensor along new dimensions.
We mechanize the semantics of our language and prove rewriting rules about $\einsum$ and $\expand$.
Our contributions are as follows.
\begin{enumerate}
\item We mechanize a language with named tensors, $\einsum$, and $\expand$, as well as its denotational semantics.
\item We show how the language we mechanize preserves the expressivity of a language with unnamed tensors.
\item We prove rewriting rules about $\einsum$ and $\expand$.
\item We %distinguish dimension names from conventional lambda-calculus variables, and
identify key definitions and properties needed to mechanize named tensors and their $\einsum$.
\item We carry out our mechanization in Lean and partially in Agda, demonstrating that our technique is not limited to a specific proof assistant.
\end{enumerate}

In \cref{section:examples}, we give motivating examples of tensor operations.
In \cref{section:syntax}, we define a language of unnamed tensors and mechanize a language of named tensors.
In \cref{section:fgg-denot}, we give the denotational semantics and demonstrate that it is preserved by a translation from the unnamed tensor language to the named tensor language.
In \cref{section:indices}, we describe the definitions and properties of named shapes and indices that underpin our mechanization.
In \cref{section:opt}, we state and prove the rewriting rules about $\einsum$ and $\expand$.
We conclude with related work in \cref{section:related}.

% section introduction

% overview: problem & solution
% not the correct one here
% the problem: opt to be proved, and the difficulty in proving it
% maybe the unnamed tensor version (at least syntax?) should be here?
% the unnamed tensor language (some old contents can be used?)
% the named tensor language (how to write this language down?)
% a compiler (maybe one way? uncompile it?) and its correctness
% the proof: tradeoff of encoding (mainly the index variable part)
\section{Motivating Examples}
\label{section:examples}

We give a series of examples to illustrate the tensor operations we mechanize and their interactions.
For each example, we write down an equation between the operation and the expected result.
If necessary, we write down an intermediate step to explain the operation in the equation.

%\subsection{Terminology}

In our terms, an \emph{unnamed tensor} is an array whose \emph{shape} is a list of natural numbers $(u_{1},\dots,u_{n})$.
The length~$n$ of this list is the \emph{rank} of the tensor.
Each number~$u_{j}$ in the list is the \emph{size} of a \emph{dimension}.
An \emph{index} into the tensor is a list $(i_{1},\dots,i_{n})$ of numbers such that $0 \le i_{j}<u_{j}$.
A rank-0 tensor is a \emph{scalar}.
A rank-1 tensor is a \emph{vector}.
A rank-2 tensor is a \emph{matrix}.

\subsection{$\einsum$}
\label{section:eg-einsum}

The $\einsum$ operation expresses a wide range of multiplications in linear algebra \cite{wenigSyntaxSemanticsEinsum2025,breuerEinsumTreesAbstraction2025}: inner product of two vectors, outer product of two vectors, matrix product, matrix transposition, Hadamard product, Kronecker product, etc.

This operation takes a list of tensors, specifies a name for each of their dimensions, and picks output dimensions from the specified names.
We use the form $\einsum(E[\vec{x}],\dotsc \to \vec{z})$, where each input $E[\vec{x}]$ pairs an unnamed tensor~$E$ with a corresponding list of dimension names $\vec{x}$, and $\vec{z}$ is the list of dimension names corresponding to the output.
For example, scaling a vector $[3\,2]$ by $5$ can be written as:
\begin{equation}
\einsum([3\,2][\ivx], 5[] \to \ivx) = [15\,10]
\end{equation}
Here, the vector $[3\,2]$ has one dimension, which we name~$x$.
The scalar $5$ has no dimension to name.

We can also express the inner product between $[3\,2]$ and $[20\,10]$ by eliminating output dimension names:
\begin{multline}
  \einsum([3\,2][\ivx],[20\,10][\ivx] \to ) \\=
  \sum\nolimits_x [3\,2]_{x} \cdot [20\,10]_{x} = 80
\end{multline}
Here we use the subscript~$_x$ to mean indexing a tensor by $x\in\{0,1\}$.
The dimension~$x$ is eliminated from the output and thus summed out.
If we keep the output dimension $x$ instead, we get elementwise multiplication:
\begin{align}
  \einsum([3\,2][\ivx],[20\,10][\ivx] \to \ivx) = [60\,20]
\end{align}

The output of $\einsum$ cannot introduce new dimensions.
For example, the $\einsum$ below is invalid, since its output uses a name $y$ that doesn't occur as any input dimension:
\begin{equation}
  \einsum([3\,2][\ivx],[20\,10][\ivx] \to {\color{red}{\ivy}})
\end{equation}
If we use $y$ as the dimension name of $[20\,10]$, we get a valid $\einsum$ operation:
\begin{multline}
  \einsum([3\,2][\ivx],[20\,10][\ivy] \to \ivy) \\=
  \Bigl[\sum\nolimits_x 20 \cdot [3\,2]_{x} \,,\quad \sum\nolimits_x 10 \cdot [3\,2]_{x}\Bigr] = [100\,50]
\end{multline}
The dimension name $x$ is again eliminated from the output, so elements along this dimension are summed up.
In contrast, when we use both $x$ and $y$ in the output dimensions, we get the outer product of $[3\,2]$ and $[20\,10]$:
\begin{equation}
  \label{e:einsum-42-xy}
  \einsum([3\,2][\ivx],[20\,10][\ivy] \to \ivx\ivy) = \bigl[\begin{smallmatrix}60&30\\40&20\end{smallmatrix}\bigr]
\end{equation}

Returning to the opening example, given any matrix~$M$, the expression $\einsum(M[ij], M[kj] \to ik)$ computes a square matrix $MM^T$, of which each element $[i,k]$ is the sum-of-product $\sum_j M[i,j]M[k,j]$.
The first dimension of~$M$ is named both $i$ and $k$ by the $\einsum$ operation, even though $M$ itself does not come with dimension names.

\subsection{$\expand$}
\label{section:eg-expand}

The $\expand$ operation adds dimensions to a tensor by repeating its contents.
It~is implemented either by making multiple elements point to the same memory location \cite{Ansel_PyTorch_2_Faster_2024,harris2020array} or by copying existing elements to new memory locations \cite{tensorflow2015-whitepaper}.

This operation takes a tensor, specifies a distinct name for each of its dimensions, and adds new dimensions to it.
We use the form $\expand(E[\vec{x}] \to \vec{z})$, where $E$ is an unnamed tensor, and $\vec{x}, \vec{z}$ are the input and output lists of dimension names.
For example, a~scalar~$4$ can be $\expand$ed to a vector:
\begin{equation}
\expand(4[]\to \ivx*) = [4\,4]
\end{equation}
The size~$2$ of the new dimension, indicated by the superscript $^{(2)}$, is actually part of every dimension name but often elided.

A vector can be $\expand$ed to a matrix in various ways:
\begin{align}
  \label{e:expand-42-1}
  \expand([3\,2][\ivx*] \to \ivx*\ivy*) &= [\begin{smallmatrix}3&3\\2&2\end{smallmatrix}]
  \\
  \label{e:expand-42-2}
  \expand([3\,2][\ivx*] \to \ivz*\ivx*) &= [\begin{smallmatrix}3&2\\3&2\end{smallmatrix}]
  \\
  \label{e:expand-42-3}
  \expand([3\,2][\ivx*] \to \ivx*\ivw*) &= [\begin{smallmatrix}3&3&3\\2&2&2\end{smallmatrix}]
\end{align}
The different results in \labelcref{e:expand-42-1} and \labelcref{e:expand-42-2} show that order matters.

The expanded matrix \labelcref{e:expand-42-3} can be further expanded:
\begin{multline}
  \expand(\expand([3\,2][\ivx] \to \ivx\ivw)[\ivx\ivw] \to \ivz\ivx\ivw) \\=
  \label{e:expand-42-4}
  \bigl[[\begin{smallmatrix}3&3&3\\2&2&2\end{smallmatrix}]\,[\begin{smallmatrix}3&3&3\\2&2&2\end{smallmatrix}]\bigr]
\end{multline}
We can merge the two nested $\expand$s above into one:
\begin{align}
  \label{e:expand-42-5}
  \expand([3\,2][\ivx] \to \ivz\ivx\ivw) =
  \bigl[[\begin{smallmatrix}3&3&3\\2&2&2\end{smallmatrix}]\,[\begin{smallmatrix}3&3&3\\2&2&2\end{smallmatrix}]\bigr]
\end{align}

\subsection{Nesting $\einsum$ and $\expand$}

Let $E = \bigl[\begin{smallmatrix}60&30\\40&20\end{smallmatrix}\bigr]$ be the result of \labelcref{e:einsum-42-xy}.
We can use $\einsum$ to combine it with the result of~\labelcref{e:expand-42-5}:
\begin{multline}
  \label{eq:expand-einsum-1}
  \einsum(\expand([3\,2][\ivx] \mathbin\to \ivz\ivx\ivw)[\ivz\ivx\ivw],E[\ivx\ivy] \to \ivz\ivx\ivw\ivy)
\\=
  \biggl[
  \biggl[
  \Bigl[
  \begin{smallmatrix}
    180&90\\
    180&90\\
    180&90
  \end{smallmatrix}
  \Bigr]
  \Bigl[
  \begin{smallmatrix}
    80&40\\
    80&40\\
    80&40
  \end{smallmatrix}
  \Bigr]
  \biggr]\quad
  \biggl[
  \Bigl[
  \begin{smallmatrix}
    180&90\\
    180&90\\
    180&90
  \end{smallmatrix}
  \Bigr]
  \Bigl[
  \begin{smallmatrix}
    80&40\\
    80&40\\
    80&40
  \end{smallmatrix}
  \Bigr]
  \biggr]
  \biggr]
\end{multline}
This result, a~tensor of shape $(2,2,3,2)$, can also be obtained by applying $\einsum$ and $\expand$ in the opposite order:
\begin{equation}
  \label{eq:expand-einsum-2}
  \expand(\einsum([3\,2][\ivx],E[\ivx\ivy] \to \ivx\ivy)[\ivx\ivy] \to \ivz\ivx\ivw\ivy)
\end{equation}
The reordering is possible because the $\einsum$ in \labelcref{eq:expand-einsum-1} doesn't need any summation.
We say this $\einsum$ is \emph{summation-free}.
This swap makes $\einsum$ operate on a smaller tensor, which improves the performance of the whole operation.

The general case of $\einsum$ computes a sum product.
To illustrate this, change the output dimensions in \labelcref{eq:expand-einsum-1} to $\ivx\ivw$:
\begin{multline}
  \label{eq:expand-einsum-3}
  \einsum(\expand([3\,2][\ivx] \mathbin\to \ivz\ivx\ivw)[\ivz\ivx\ivw], E[\ivx\ivy] \to \ivx\ivw)
\\=
  \bigl[
  \begin{smallmatrix}
    540&540&540\\
    240&240&240\\
  \end{smallmatrix}
  \bigr]
\end{multline}
Again let us apply $\einsum$ and $\expand$ in the opposite order:
\begin{multline}
  \label{eq:expand-einsum-4}
  \expand(\einsum([3\,2][\ivx],E[\ivx\ivy] \to \ivx)[\ivx] \to \ivx\ivw)
\\=
  \bigl[
  \begin{smallmatrix}
    270&270&270\\
    120&120&120\\
  \end{smallmatrix}
  \bigr]
\end{multline}
The results in \labelcref{eq:expand-einsum-3} and \labelcref{eq:expand-einsum-4} differ by a factor of~2, which is the size of the $\ivz$ dimension.
We need to make up for the factor to preserve the result when swapping $\einsum$ with $\expand$.
Based on this idea, we prove the general commutativity rule between $\einsum$ and $\expand$.

\section{The Syntax of a Tensor Language}
\label{section:syntax}

We give the syntax of our tensor language, in two versions.
In~the first version, expressions denote tensors with unnamed dimensions, as conceived usually in tensor libraries and illustrated in \cref{section:examples}.
In~the second version, expressions denote tensors with named dimensions.
We mechanize the second version, and we specify the necessary conditions to guarantee well-formedness in a dependently typed proof assistant.
To show the correspondence between the two versions, we give a translation in \cref{section:syntax-translate}.

\subsection{An Unnamed Tensor Language}
\label{section:unnamed-syntax}
\begin{figure}
$
\begin{array}{@{}l@{}l@{}}
    \text{Tensor expressions} \quad %&& (\data{Tensor}{}{})
    E,F ::={} & \alpha \wmid 0^{(s,\ldots)} \wmid n \cdot E \wmid E+E \wmid \\[\jot]
    &\bind{t}{E}{E} \wmid t \wmid \\[\jot]
    &\expand(E[\vec{x}] \to \vec{z}) \wmid \\[\jot]
    &\einsum(E[\vec{x}],\dotsc \to \vec{z}) \\[\jot] % \wmid \\
    % &&&\scatter(E[\vec{x}] \to e\dotsm) \\
    \text{Tensor variables} %&& (\data{TVar}{}{})
    & t \\[\jot]
    \text{Index variable sequences} %&& (\data{IVars}{}{})
    & \vec{x},\vec{y},\vec{z} ::= x^{(s)}\dotsm \\[\jot]
    % &\text{Axis expressions} && (\data{Axis}{}{}) & e &::={} && x^{(s)} \wmid () \wmid s\oplus e\oplus s \wmid e\otimes e \\
    \text{Index variables} %&& (\data{IVar}{}{})
    & x^{(s)},y^{(s)},z^{(s)} \\[\jot]
    \text{Scalar} & \alpha \in \mathbb{S}\\[\jot]
    \text{Upper bounds} & s \in \mathbb{N}\\[\jot]
    \text{Scale} & n \in \mathbb{N}
\end{array}
$
\caption{The syntax of an unnamed tensor language}
\label{fig:fgg-syntax}
\end{figure}

The syntax of an unnamed tensor language is shown in \cref{fig:fgg-syntax}.
A tensor expression can be a literal scalar~$\alpha$, a zero tensor of a specified shape $(s,\ldots)$, a tensor scaled by a natural number, a sum of two tensors, a tensor variable, a binding, an $\expand$, or an $\einsum$.
To specify the input and output of $\expand$ and $\einsum$, we use \emph{index variable sequences}, which are lists of \emph{index variables}.
Below we explain each construct in turn.

\subsubsection{Constants}
Constant tensors include scalars and zero tensors.
We require that a scalar range over a semiring $\mathbb{S}$.
A scalar always has the shape $()$.
Therefore, the scalar $0$ can either be written as a scalar $0$ or a zero tensor $0^{()}$.

A zero tensor is specified by a shape $(s,\ldots)$, a list of natural numbers.
All of its elements are $0$ in the semiring.
A zero tensor can also be expressed by $\expand$ing a scalar:
\begin{equation}
  0^{(s_{1},\dots,s_{n})} = \expand(0[] \to \ivar{x_{1}}{s_{1}}\cdots\ivar{x_{n}}{s_{n}})
\end{equation}
%where $x_{1},\cdots,x_{n}$ are fresh names.

\subsubsection{Scaling}
A scaling expression $n \cdot E$ multiplies each element in $E$ by the natural number $n$.
Here multiplication means repeated addition, so we can express it by using $+$ repeatedly.
%: \[n \cdot E = \underbrace{E + \cdots + E}_{n~\text{times}}.\]
We can also express it using $\einsum$ and $\expand$:
\begin{equation}
    n \cdot E = \einsum(\expand(1[] \to \ivar{x}{n})[\ivar{x}{n}], E[\vec{y}] \to \vec{y})
\end{equation}
where $x$ is a name that doesn't occur in~$\vec{y}$.
%We provide scaling for convenience.

\subsubsection{Addition}
The expression $E + F$ performs semiring addition elementwise.
$E$~and $F$ must have the same shape.

\subsubsection{Tensor variables and $\bindop$}
A \emph{tensor variable}~$t$ is introduced by a $\bindop$ expression.
As explained in \cref{section:fgg-denote}, a~tensor variable denotes an unnamed tensor, whose shape is a list of natural numbers.
To~use the variable, we simply name~it, but we assume that its shape is correct and known.
For example, in a tensor expression $t_{1} + t_{2}$, we~assume that $t_{1}$ and~$t_{2}$ have the same shape, which can be looked up from the $\bindop$ expressions where $t_{1}$ and~$t_{2}$ are bound.

\subsubsection{$\expand$}
As shown in \cref{section:eg-expand}, an $\expand$ expression transforms a tensor to a higher rank by repeating its contents.
In~the general form $\expand(E[\vec{x}] \to \vec{z})$, the input index variables in~$\vec{x}$ must be distinct, the output index variables in~$\vec{z}$ must be distinct, and every index variable occurring in $\vec{x}$ must also occur in~$\vec{z}$.
\begin{comment}
In addition, the upper bounds in $\vec{x}$ must match the shape of~$E$.
For example, suppose we want to expand the result $[\begin{smallmatrix}3&3&3\\2&2&2\end{smallmatrix}]$ of \cref{e:expand-42-3}, a tensor of shape $(2, 3)$.
We name this tensor as $t'$.
We can write
\begin{equation}
    \expand(t'[\ivx*\ivw*] \to \ivz*\ivx*\ivw*),
\end{equation}
but not
\begin{equation}
    \expand(t'[{\color{red}{\ivw*\ivx*}}] \to \ivz*\ivx*\ivw*),
\end{equation}
because the upper bounds of the index variable sequence $\ivw*\ivx*$ are $(3, 2)$, rather than $(2, 3)$.
\end{comment}

\subsubsection{$\einsum$}
As shown in \cref{section:eg-einsum}, $\einsum$ defines each output element as a sum of products of input elements.
To~specify which dimensions to sum over and which to multiply together, $\einsum$ uses index variables to relate the input tensors' dimensions to the output tensor's dimensions.

In~the general form $\einsum(E[\vec{x}],\dotsc \to \vec{z})$,
% TODO: maybe remove this sentence when repeated input dimension names are supported.
variables in each input variable sequence $\vec{x}$ must be distinct, but one index variable can occur in multiple input index variable sequences.
The output index variables in~$\vec{z}$, if any, must be distinct and must each appear at least once among the input $\vec{x}$ sequences.
These restrictions rule out the problematic sequence $ii$ in \cref{section:intro}.

\subsubsection{Index variables}
An \emph{index variable} $\ivar{x}{s}$ is a name $x$ together with an \emph{upper bound}~$s$.
It tells $\expand$ and $\einsum$ which tensor dimensions to operate~on.
For instance, in $\expand([3\,2][\ivx*] \to \ivx*\ivw*)$ from \labelcref{e:expand-42-3}, the variable~$\ivx*$ specifies the only dimension of the input tensor $[3\,2]$ to be of size~$2$ and names it as~$x$.
The variable $\ivw*$ specifies where to add the new dimension and its size~$3$.
It is an error to operate on a tensor whose shape does not match the upper bound of the index variable, as in $\expand([3\,2][\ivw*] \to \ivx*\ivw*)$.

The index variable~$x^{(s)}$ iterates over integers between $0$ and $s-1$.
We write $|x^{(s)}|$ for the upper bound~$s$.

We write $\vec{x}$ or $\vec{y}$ or $\vec{z}$ for an \emph{index variable sequence}.
We notate concatenation of sequences by juxtaposition:~$\vec{x}\vec{y}$.
We define the union sequence $\vec{x}\cup\vec{y}$ to be the concatenation $\vec{x}\vec{y}$ with duplicates removed (leaving the first occurrence of each index variable, say, so $xy \cup xzx = xyz$).
We write $\#\vec{x}$ for the number of index variables in $\vec{x}$.
%TODO: (where) do we need to notate the number of index variables in an index variable sequence?
We write $|\vec{x}|$ for the list of upper bounds of the index variables in~$\vec{x}$.

Index variables in different operations are interpreted in separate scopes.
For instance, the nested $\expand$ from \cref{e:expand-42-4} can be rewritten as
\begin{equation}
\label{e:expand-rename}
  \expand(\expand([3\,2][\ivx] \to \ivx\ivw)[\ivz\ivw] \to \ivar*{\underline{x}}{2}\ivz\ivw).
\end{equation}
The underlined $\underline{x}$ is interpreted as a new dimension added to the input of the outer $\expand$, rather than referring to the one without underline from the inner $\expand$.

Thus, when nesting one $\expand$ inside another, the index variable sequences for the inner output and the outer input may use different names.
In~\labelcref{e:expand-rename} for example, the inner output $\ivx*\ivw*$ and the outer input $\ivz*\ivw*$ specify the same shape $|\ivx*\ivw*| = |\ivz*\ivw*| = (2,3)
$ but have different names.
For two index variable sequences $\vec{x}$ and~$\vec{y}$, as long as $|\vec{x}| = |\vec{y}|$, we can rename index variables between $\vec{x}$ and~$\vec{y}$.
This idea is used in \cref{section:syntax-translate} to translate the unnamed tensor language to a named version and eliminate alpha-conversion.

\begin{figure}
\begin{lstlisting}
inductive TExpr {S : Type} [Semiring S] :
  Context → Is → Type where
  | scalar {Γ : Context} : S → TExpr Γ empty
  | zero {Γ : Context} : (xs : Is) → TExpr Γ xs
  | scale {Γ : Context} {xs : Is} :
    ℕ → TExpr Γ xs → TExpr Γ xs
  | add {Γ : Context} {xs ys : Is} :
    TExpr Γ xs → TExpr Γ ys → xs ≈ ys →
    TExpr Γ xs
  | let {Γ : Context} {xs : Is} :
    (dims : List IVar) → (neq : Unique dims) →
    (rhs : TExpr Γ (fromList dims neq)) →
    (body : TExpr (Γ `, (dims.map Prod.snd)) xs) →
    TExpr Γ xs
  | var {Γ : Context} :
    (dims : List IVar) → (neq : Unique dims) →
    (dims.map Prod.snd ∈ᵘ Γ) →
    TExpr Γ (fromList dims neq)
  | expand {Γ : Context} {xs zs : Is} :
    TExpr Γ xs → xs ⊑ zs → TExpr Γ zs
  | einsum {Γ : Context} {xss : List Is} {zs : Is} :
    TExprSeq Γ xss →
    (xssCompat : IssCompat xss) →
    zs ⊑ ⋃ xssCompat →
    TExpr Γ zs
\end{lstlisting}
\vspace{-5pt}
\caption{The syntax of a named tensor language}
\label{fig:named-syntax}
\vspace{-5pt}
\end{figure}

\subsection{A Named Tensor Language}
\label{section:named-syntax}

The syntax of a named tensor language in Lean is shown in \cref{fig:named-syntax}.
In this definition, a \lstinline{TExpr} is an inductive type parameterized by a value type \lstinline{S} and indexed by a \lstinline{Context} and an \lstinline{Is}.
It represents a well-formed tensor whose dimensions are named.
We require that \lstinline{S} be an instance of \lstinline{Semiring}.% and add \lstinline{Semiring S} as an implicit instance parameter.

A \lstinline{Context} stores the shapes of tensor variables in de Bruijn style.
It is defined as the following inductive data type:
\begin{lstlisting}
inductive Context : Type where
  | nil : Context
  | snoc : Context → List UpperBound → Context
infixl:55 " `, " => Context.snoc
\end{lstlisting}
As \lstinline{List UpperBound} above indicates, a shape in a \lstinline{Context} is a list of \emph{unnamed} dimension sizes.
Here \lstinline{UpperBound} is an alias for natural numbers that store a dimension size:
\begin{lstlisting}
abbrev UpperBound := ℕ
\end{lstlisting}

An \lstinline{Is} is a finite map that represents the shape of a tensor whose dimensions are named.
Therefore, we use \emph{named shape} to mean an \lstinline{Is} for \lstinline{TExpr} below.
For example, the named shape of \labelcref{e:expand-42-5} in the named tensor language can be written as
\begin{equation}
\{w \mapsto 3, x \mapsto 2, z \mapsto 2\}.
\end{equation}
We leave the details of \lstinline{Is} to \cref{section:indices}.
Because a \lstinline{TExpr} is indexed by an \lstinline{Is}, we always have the dimension names and sizes for any given tensor.
Thus, the type of \labelcref{e:expand-42-5} in the empty \lstinline{Context} \lstinline{nil} can be written as
\begin{lstlisting}
TExpr (S := ℕ) nil {w ↦ 3, x ↦ 2, z ↦ 2}
\end{lstlisting}

The constructors of \lstinline{TExpr} correspond to each case of tensor expressions in \cref{fig:fgg-syntax}.
We explain the constraints introduced by each constructor below.

\subsubsection{Constants}

The constructor \lstinline{scalar} takes a value of type \lstinline{S}.
Its named shape is the \lstinline{empty} map.
The constructor \lstinline{zero} simply takes a named shape \lstinline{xs}.

\subsubsection{Scaling}

The constructor \lstinline{scale} takes a factor \lstinline{n} and a tensor of named shape \lstinline{xs}.
The result also has shape \lstinline{xs}.

\subsubsection{Addition}

The constructor \lstinline{add} takes two tensors, of~named shapes \lstinline{xs} and~\lstinline{ys}.
Here we require that \lstinline{xs} and~\lstinline{ys} be \emph{equivalent}, written \lstinline{xs ≈ ys}: for any index variable name~$x$, if~$x$ is in~\lstinline{xs} and has the size~\lstinline{up}, then $x$ is in~\lstinline{ys} and also has the size \lstinline{up}, and vice versa.
For example,
\begin{lstlisting}
{w ↦ 3, x ↦ 2, z ↦ 2} ≈ {z ↦ 2, x ↦ 2, w ↦ 3}
\end{lstlisting}
This definition makes it possible to avoid explicit permutations between different \lstinline{Is}es.
Since \lstinline{xs} and \lstinline{ys} are equivalent, we can pick either as the result shape.
Here we choose \lstinline{xs}.

\subsubsection{Tensor variables}

The way tensor variables are introduced and used reflects the fact that, whereas a tensor denoted by an expression has named dimensions, a tensor associated with a variable has unnamed dimensions.
This distinction is that between R-values and L-values \cite{friedmanEssentialsProgrammingLanguages2008}.

Tensor variables are introduced by the constructor \lstinline{let}.
Since tensor variables are encoded with de Bruijn indices, \lstinline{let} doesn't need to take a tensor variable name to bind.
Instead, it takes a list of index variables, which are defined as a product of an index variable name and a dimension size:
\begin{lstlisting}
abbrev IVar := Id × UpperBound
\end{lstlisting}
We encode index variable names as \lstinline{String}s:
\begin{lstlisting}
abbrev Id := String
\end{lstlisting}
The next argument \lstinline{neq} is a proof that names in \lstinline{dims} have no duplicates.
We can construct an \lstinline{Is} with \lstinline{fromList dims neq}, which is the named shape of the right hand side tensor of \lstinline{let}.
The \lstinline{body} of \lstinline{let} is a tensor in a larger \lstinline{Context} \lstinline{(Γ `, (dims.map Prod.snd))} with the named shape \lstinline{xs}.

To refer to a tensor variable, we use \lstinline{var}, which takes three explicit arguments.
First, \lstinline{dims} lists the index variables with which to index the tensor.
Second, \lstinline{neq} requires the names in \lstinline{dims} to be unique.
Finally, the argument of type \lstinline{dims.map Prod.snd ∈ᵘ Γ} specifies which tensor variable to use, and requires the upper bounds in \lstinline{dims} to match.
The relation \lstinline{∈ᵘ} specifies the lookup of a de Bruijn-style variable:
\begin{lstlisting}[mathescape=true]
inductive CtxMem : List UpperBound → Context →
  Type where
  | here {Γ : Context} {dims : List UpperBound} :
    CtxMem dims (Γ `, dims)
  | there$\,${Γ : Context}$\,${dims dims'$\,$:$\,$List UpperBound}$\,$:
    CtxMem dims Γ → CtxMem dims (Γ `, dims')
notation dims " ∈ᵘ " Γ => CtxMem dims Γ
\end{lstlisting}

Note that the index variable names in \lstinline{let} and in \lstinline{var} can differ, so a tensor variable can be used multiple times with different index variable sequences.
For example, suppose $M$ is a matrix introduced by a \lstinline{let} whose \lstinline{dims} is \lstinline{[("a",2),("b",3)]}.
Then, to~compute $M M^T$ by $\einsum(M_{ij}, M_{kj} \to ik)$ as in \cref{section:intro}, we can refer to $M$ twice by \lstinline{var}, with the \lstinline{dims} \lstinline{[("i",2),("j",3)]} and \lstinline{[("k",2),("j",3)]} respectively.

\subsubsection{$\expand$}

The constructor \lstinline{expand} takes a tensor of named shape \lstinline{xs} and returns a tensor of named shape \lstinline{zs}.
We require that \lstinline{xs} be a \emph{subset} of \lstinline{zs}, written \lstinline{xs ⊑ zs}: for any index variable name $x$, if $x$ is in \lstinline{xs} and has the size \lstinline{up}, then $x$ is in \lstinline{zs} and also has the size \lstinline{up}.
For example,
\begin{lstlisting}[belowskip=0pt]
{x ↦ 2, z ↦ 2} ⊑ {z ↦ 2, x ↦ 2, w ↦ 3}
\end{lstlisting}

\subsubsection{$\einsum$}

The constructor \lstinline{einsum} takes a sequence of tensor expressions of named shapes \lstinline{xss} and returns a tensor of named shape \lstinline{zs}.
A sequence of tensor expressions is defined as an inductive data type \lstinline{TExprSeq}
%parameterized by a semiring \lstinline{S} and
indexed by a \lstinline{Context} and a list of \lstinline{Is}:
%The two constructors of \lstinline{TExprSeq} construct an empty and a non-empty sequence of well-formed tensor expressions respectively.
\begin{lstlisting}
inductive TExprSeq {S : Type} [Semiring S] :
  Context → List Is → Type where
  | nil {Γ : Context} : TExprSeq Γ []
  | cons {Γ : Context} {xs : Is} {xss : List Is} :
    TExpr Γ xs → TExprSeq Γ xss →
    TExprSeq Γ (xs :: xss)
\end{lstlisting}

We require the \lstinline{Is}es in \lstinline{xss} to be pairwise \emph{compatible}.
Two \lstinline{Is}es \lstinline{xs} and \lstinline{ys} are compatible, written \lstinline{xs ∼ ys}, if every index variable name~$x$ occurs either in neither \lstinline{xs} nor \lstinline{ys}, or in only one of \lstinline{xs} and \lstinline{ys}, or in both \lstinline{xs} and \lstinline{ys} with the same size.
So
\begin{lstlisting}
{x ↦ 2, z ↦ 2} ∼ {z ↦ 2, x ↦ 2, w ↦ 3}
\end{lstlisting}
but
\begin{lstlisting}[mathescape=true]
{x ↦ 2, z ↦ 2} $\not{\sim}$ {z ↦ 3, x ↦ 2, w ↦ 3}
\end{lstlisting}

We also require the result shape \lstinline{zs} to be a subset of the \emph{union} of \lstinline{xss}.
The union of a list of compatible \lstinline{Is}es, written \lstinline{⋃ xss}, is defined recursively using the \emph{union} of two \lstinline{Is}es \lstinline{xs} and \lstinline{ys}, written \lstinline{xs ∪ ys}:
for any index variable name $x$, if $x$ occurs in \lstinline{xs ∪ ys}, then $x$ occurs either in \lstinline{xs} or \lstinline{ys} with the same size, and vice versa.
For example,
\begin{lstlisting}[belowskip=0pt]
{x ↦ 2, y ↦ 4, z ↦ 2} ∪ {z ↦ 2, x ↦ 2, w ↦ 3}
≈ {z ↦ 2, y ↦ 4, x ↦ 2, w ↦ 3}
\end{lstlisting}
%With these constraints, it is guaranteed that an $\einsum$ expression is always well-formed.

\section{Denotational Semantics}
\label{section:fgg-denot}

In this section, we give the denotational semantics of the two versions of our tensor language.
The semantics of the unnamed version is given in \cref{section:fgg-denote}.
The semantics of the named version is given in \cref{section:named-denote}.
We give a semantics-preserving translation between the unnamed version and the named version in \cref{section:syntax-translate}.

\subsection{The Semantics of the Unnamed Tensor Language}
\label{section:fgg-denote}
\begin{figure}
\begin{minipage}{\columnwidth}
\centering
\begin{alignat}{2}
\label{e:fgg-scalar}
    &\llbracket \alpha \rrbracket_\gamma [()] &&=\alpha
\\
\label{e:fgg-zero}
    &\llbracket 0^{(s,\dotsc)} \rrbracket_\gamma [I] &&=0
\\
\label{e:fgg-scale}
    &\llbracket n \cdot E \rrbracket_\gamma [I] &&=n \cdot \llbracket E \rrbracket_\gamma [I]
\\
\label{e:fgg-add}
    &\llbracket E_1 + E_2 \rrbracket_\gamma [I] &&=\llbracket E_1 \rrbracket_\gamma [I] + \llbracket E_2 \rrbracket_\gamma [I]
\\
\label{e:fgg-let}
    &\llbracket \bind{t}{E_{1}}{E_{2}} \rrbracket_\gamma [I] &&= \llbracket {E_{2}} \rrbracket_{\gamma;t,\llbracket {E_{1}} \rrbracket_{\gamma}} [I]
\\
\label{e:fgg-var}
    &\llbracket t \rrbracket_\gamma [I] &&=\gamma(t) [I]
\\
\label{e:fgg-expand}
    &\llbracket \expand(E[\vec{x}] \to \vec{z}) \rrbracket_\gamma [I]
    &&=\llbracket E \rrbracket_\gamma [\{\vec{z}\mapsto I\}(\vec{x})]
\\\notag
    &\llbracket \einsum(E_1[\vec{x}_1],\dotsc,E_k[\vec{x}_k] \to \vec{z}) \rrbracket_\gamma [I]
\hidewidth\\\label{e:fgg-einsum}
    &\quad=\sum_{\delta\in\dom{\bigcup_{i=1}^k\vec{x}_i\setminus\vec{z}}} \prod\nolimits_{i=1}^k \llbracket E_i \rrbracket_\gamma [(\delta\cup\{\vec{z}\mapsto I\})(\vec{x}_i)]
\hidewidth
% \\
% \label{e:fgg-scatter}
%     &\llbracket \scatter(E[\vec{x}] \to e\dotsm) \rrbracket_\gamma [I]
%     &&=\sum\nolimits_{\begin{subarray}{l}\delta\in\dom{\vec{x}} \\ \delta(e\dotsm) = I\end{subarray}} \llbracket E \rrbracket_\gamma [\delta(\vec{x})]
\end{alignat}
\end{minipage}
\caption{The denotations of unnamed tensor expressions}
\label{fig:fgg-denote}
\end{figure}

The denotational semantics of the unnamed tensor language is shown in \cref{fig:fgg-denote}.
The denotation $\denot{E}_{\gamma}$ of an unnamed tensor expression $E$ in an \emph{environment} $\gamma$ is an unnamed tensor.
An environment is a function from tensor variable names to unnamed tensors.
An unnamed tensor is a function that takes an index~$I$ and returns a value from the semiring~$\mathbb{S}$.
An index is a list of natural numbers.

The first four cases of \cref{fig:fgg-denote} are simple.
\Cref{e:fgg-scalar} says that a literal scalar has value $\alpha$.
\Cref{e:fgg-zero} says that a zero tensor has value 0 for any index $I$.
\Cref{e:fgg-scale} says that the denotation of $n \cdot E$ is the denotation of $E$ multiplied by $n$.
\Cref{e:fgg-add} requires that the two tensors have the same shape and adds the elements at the location specified by $I$.

\Cref{e:fgg-let} says that we get the denotation of the right hand side $E_{1}$ of the $\bindop$ expression in $\gamma$, extend $\gamma$ with $t$ bound to the denotation of $E_{1}$, and then get the denotation of $E_{2}$ in the extended environment.
\Cref{e:fgg-var} says that we look up a tensor variable $t$ in $\gamma$ and pass $I$ to its tensor value.

To define the denotation of $\expand$ and $\einsum$, we first introduce \emph{named indices}.
A \emph{named index} is a function $\delta$ from some index variables to non-negative integers, such that $0 \le \delta(x^{(s)}) < s$ for each $x^{(s)}$ in the domain of~$\delta$.
Let $\vec{x} = x_1^{(s_1)} \dotsm x_r^{(s_r)}$ be a sequence of $r$ index variables, not necessarily distinct.
If~$\delta$ is a named index whose domain includes~$\vec{x}$, then we write $\delta(\vec{x})$ for the index $\bigl( \delta(x_1), \dotsc, \delta(x_r) \bigr)$, which can be used to index a tensor of shape $(s_1,\dotsc,s_r)$.
Conversely, if~$I$ is $(i_1,\dotsc,i_r)$, then we write $\{\vec{x}\mapsto I\}$ for the named index $\{ x_1\mapsto i_1, \dotsc, x_r\mapsto i_r \}$; here we do require the index variables in~$\vec{x}$ to be distinct.
We also define the set $\dom{\vec{x}}$ to consist of all named indices whose domain is~$\vec{x}$.%, and the set $\tups{\vec{x}}$ to consist of all indices $(i_1,\dotsc,i_r)$ that satisfy $0\le i_1<s_1, \dotsc, 0\le i_r<s_r$.

\Cref{e:fgg-expand} says, to~look up an element of an $\expand$ed tensor, we~look up the corresponding element of the original tensor, by~removing from the index the new dimensions introduced by $\expand$.
For instance, the element of $\llbracket \expand(E[x] \to xy) \rrbracket_\gamma$ at $(i,j)$ is the element of $\llbracket E \rrbracket_\gamma$ at~$(i)$.

\Cref{e:fgg-einsum} defines the denotation of an $\einsum$ expression by iterating over all indices and summing out those index variables that do not appear in the output~$\vec{z}$. (Recall that our syntax requires $\vec{z}$ to be a subset of $\bigcup_{i=1}^k\vec{x}_i$.)
For instance, multiplying two matrices $E_1$ and~$E_2$ is denoted by $\einsum(E_1[xy],E_2[yz] \to xz)$, because \labelcref{e:fgg-einsum} amounts to
\begin{multline}
  \llbracket\einsum(E_1[xy],E_2[yz] \to xz)\rrbracket_\gamma[(i,k)]
  \\=\sum\nolimits_j \llbracket E_1\rrbracket_\gamma[(i,j)] \llbracket E_2\rrbracket_\gamma[(j,k)].
\end{multline}
\begin{comment}
In the special case where $\vec{z} = \bigcup_{i=1}^k\vec{x}_i$, we say that the $\einsum$ is \emph{summation-free}:
\begin{multline}
\label{e:fgg-einsum-summation-free}
  \llbracket \einsum(E_1[\vec{x}_1],\dotsc,E_k[\vec{x}_k] \to \vec{z}) \rrbracket_\gamma [I] \\=
  \prod\nolimits_{i=1}^k \llbracket E_i \rrbracket_\gamma [\{\vec{z} \mapsto I\}(\vec{x}_i)]
\end{multline}
Otherwise, we say $\einsum$ is \emph{non-summation-free}.
The right hand side of~\labelcref{e:fgg-einsum-summation-free} does not quite reduce to the right hand side of~\labelcref{e:fgg-expand} when $k=1$, because $\expand$ allows $\vec{z}$ to contain index variables not present in~$\vec{x}$.
\end{comment}

\subsection{The Semantics of the Named Tensor Language}
\label{section:named-denote}

The denotational semantics of the named tensor language is shown in \cref{fig:named-denote}.
\begin{figure}
\begin{lstlisting}
def denote {S : Type} [Semiring S]
  {Γ : Context} {xs : Is}
  (t : TExpr (S := S) Γ xs)
  (env : Env (S := S) Γ)
  (idx : NIndex xs) : S :=
  match t with
  | .scalar r => r
  | .zero _ => 0
  | .scale n t => n * denote t env idx
  | .add t₁ t₂ xs_ys =>
    (denote t₁ env idx) +
    (denote t₂ env (idx →ᵢ xs_ys))
  | .let dims neq rhs body =>
    let rhsNamed := denote rhs env
    let rhsUnnamed idx' :=
      rhsNamed (NIndex.fromUnnamed dims neq idx')
    denote body (env , rhsUnnamed) idx
  | .var dims neq v =>
    lookupUnnamed v env
      (NIndex.toUnnamed dims neq idx)
  | .expand t xs_zs => denote t env (idx ↓ᵢ xs_zs)
  | .einsum (zs := zs) ts xssCompat zs_xss =>
    have zs_disjoint : zs ⊥ (⋃ xssCompat ∖ zs) :=
      minusIs_disjoint
    have zs_union :
      zs ∪ (⋃ xssCompat ∖ zs) ≈ ⋃ xssCompat :=
      minusIs_union_eq zs_xss
    sum (⋃ xssCompat ∖ zs)
      (fun idx' =>
        denotes ts xssCompat
          (IsSubset_refl (xs := ⋃ xssCompat)) env
          ((∪ᵢ zs_disjoint idx idx') →ᵢ zs_union))
\end{lstlisting}
\caption{The denotations of named tensor expressions}
\label{fig:named-denote}
\vspace{-15pt}
\end{figure}
In this definition, the function \lstinline{denote} is parameterized by a \lstinline{Semiring S}, a \lstinline{Context} $\Gamma$, and a named shape \lstinline{xs}.
It then takes a tensor expression \lstinline{t}, an environment \lstinline{env}, and a \emph{named index} \lstinline{idx} of type \lstinline{NIndex xs}, before returning a value of type~\lstinline{S}.
Therefore, the denotation of a named tensor expression is a \emph{named tensor}, which is a function that takes a named index and returns a value from the semiring~\lstinline{S}.
We explain the arguments and definition of \lstinline{denote} below.

\subsubsection{Environment}
The environment \lstinline{env} has type \lstinline{Env (S := S) Γ}, which depends on the \lstinline{Context}~\lstinline{Γ}.
Each unnamed shape \lstinline{dims} in~\lstinline{Γ} calls for an unnamed tensor in~\lstinline{env}, whose type~\lstinline{UnnamedTensor dims} we define shortly.
\begin{lstlisting}
def Env {S : Type} [Semiring S] : Context → Type
  | .nil => Unit
  | Γ `, dims =>
    Env (S := S) Γ × UnnamedTensor (S := S) dims
\end{lstlisting}
An \lstinline{UnnamedTensor} of unnamed shape \lstinline{dims} is a function from an index of type \lstinline{UIndex dims} to a value of type \lstinline{S}:
\begin{lstlisting}
abbrev UnnamedTensor {S : Type} [Semiring S]
  (dims : List UpperBound) :=
  UIndex dims → S
\end{lstlisting}
An (unnamed) index, or \lstinline{UIndex}, of shape \lstinline{dims} is a sequence of natural numbers, each bounded by the corresponding element \lstinline{d} of \lstinline{dims}:
\begin{lstlisting}[belowskip=0pt]
def UIndex (dims : List UpperBound) : Type :=
  match dims with
  | [] => Unit
  | d :: ds => Fin d × UIndex ds
\end{lstlisting}

\subsubsection{Named indices}

Conceptually, a named index is a finite map from names to numbers that are bounded by a named shape.
For example, if \lstinline{xs} is the named shape \lstinline|{z ↦ 2, x ↦ 2, w ↦ 3}|, then the argument \lstinline{idx : NIndex xs} in \cref{fig:named-denote} can be the named index \lstinline|{z ↦ 1, x ↦ 0, w ↦ 2}ᵢ|.
The type \lstinline{NIndex xs} is defined as a structure with a field \lstinline{get}.
This field is a function that takes an index variable name \lstinline{x} and a dimension size \lstinline{u}.
If \lstinline{get} looks up \lstinline{x} in \lstinline{xs} and finds the value \lstinline{u}, it returns a number bounded by~\lstinline{u}:
\begin{lstlisting}
structure NIndex (xs : Is) where
  get : ∀ {x : Id} {u : UpperBound},
    lookup xs x = some u → Fin u
\end{lstlisting}
Here, \lstinline{lookup} is a function that takes a named shape \lstinline{xs} and an index variable name \lstinline{x}.
It returns an \lstinline{Option UpperBound} to indicate if \lstinline{x} is found in \lstinline{xs}:
\begin{lstlisting}[belowskip=0pt]
def lookup (xs : Is) (x : Id) : Option UpperBound
\end{lstlisting}

\subsubsection{The definition of \lstinline{denote}}
The definition of \lstinline{denote} in \cref{fig:named-denote} closely follows the definition of $\denot{}$ in \cref{fig:fgg-denote}.

The first three cases are simple.
A scalar always has the value \lstinline{r}.
A zero tensor always has the value 0.
The denotation of a scaled tensor is the denotation of the tensor multiplied by a natural number.

The denotation of \lstinline{add t₁ t₂ xs_ys} at a named index \lstinline{idx} depends on the denotations of \lstinline{t₁} and \lstinline{t₂}.
For \lstinline{t₁}, we simply use the index \lstinline{idx}, which is of \lstinline{t₁}'s shape.
For \lstinline{t₂}, since its shape is equivalent to \lstinline{t₁}'s, we use \lstinline{→ᵢ} to convert \lstinline{idx} from \lstinline{t₁}'s shape to \lstinline{t₂}'s shape.
Since this conversion is based on the equivalence of two shapes, we do not need to explicitly define permutation between indices.

The key to tensor variable binding and reference is a pair of functions that converts between a named index and an unnamed index, \lstinline{NIndex.fromUnnamed} and \lstinline{NIndex.toUnnamed}.
For example, given an unnamed shape $(2, 2, 3)$ and a corresponding unnamed index $(1, 0, 2)$, if we know the three dimensions are named as $z, x, w$, then we can get a named index \lstinline|{z ↦ 1, x ↦ 0, w ↦ 2}ᵢ|.
Similarly, given a named shape \lstinline|{w ↦ 3, x ↦ 2, z ↦ 2}| and a named index $\{w \mapsto 2, x \mapsto 0, z \mapsto 1\}_{i}$, if we know the dimension order is $(z, x, w)$, we can construct an unnamed index \lstinline|(1, 0, 2)|.

The \lstinline{let} case first gets the denotation of \lstinline{rhs} in \lstinline{env}, which is a named tensor \lstinline{rhsNamed}.
Then it constructs an unnamed tensor \lstinline{rhsUnnamed} from \lstinline{rhsNamed} by converting an unnamed index \lstinline{idx'} to a named index with \lstinline{NIndex.fromUnnamed}.
\begin{comment}
That~is,
\begin{lstlisting}
rhsUnnamed = rhsNamed ∘ (NIndex.fromUnnamed dims neq)
\end{lstlisting}
\end{comment}
Finally, it evaluates \lstinline{body} in \lstinline{env} extended with \lstinline{rhsUnnamed}.

The \lstinline{var} case first invokes \lstinline{lookupUnnamed} to retrieve the value stored for \lstinline{v} in \lstinline{env}, which is an unnamed tensor.
Then it converts the given named index \lstinline{idx} to unnamed with \lstinline{NIndex.toUnnamed}.
Finally, it passes the unnamed index to the unnamed tensor.

The \lstinline{expand} case uses \lstinline{↓ᵢ} to remove from the named index \lstinline{idx} the dimensions that are not in \lstinline{t₁}'s shape, then gets the denotation of \lstinline{t₁}.

Finally, the \lstinline{einsum} case performs summation using a helper function \lstinline{sum}, which has the following type:
\begin{lstlisting}
def sum {S : Type} [Semiring S]
  (xs : Is) (f : NIndex xs → S) : S
\end{lstlisting}
Given a named shape \lstinline{xs}, it enumerates all possible indices of \lstinline{xs} and passes them to \lstinline{f}, a function that takes a named index of \lstinline{xs} and returns a value.
Then it sums up all values returned by \lstinline{f} together.

In \lstinline{denote}, the first argument to \lstinline{sum} is the named shape returned by a function \lstinline{∖}, which removes \lstinline{zs} from \lstinline{⋃ xss}, the union of all input shapes.
The second argument to \lstinline{sum} is a function that computes the product of all input tensors, using the helper function \lstinline{denotes}, whose type is below:
\begin{lstlisting}
def denotes {S : Type} [Semiring S]
  {Γ : Context} {xss : List Is} {ys : Is}
  (ts : TExprSeq (S := S) Γ xss)
  (xssCompat : IssCompat xss)
  (h : ⋃ xssCompat ⊑ ys)
  (env : Env (S := S) Γ) (idx : NIndex ys) : S
\end{lstlisting}
The function \lstinline{denotes} takes a sequence of tensor expressions \lstinline{ts} of named shapes \lstinline{xss}, a named index \lstinline{idx} of shape \lstinline{ys}, and a proof~\lstinline{h} that the union of \lstinline{xss} is a subset of \lstinline{ys}.
When doing recursion on \lstinline{ts}, we keep the invariant that the union of \lstinline{xss} is a subset of \lstinline{ys}.
For each tensor in the sequence, it gets the denotation of the tensor at the subset of \lstinline{idx} that matches the tensor's shape.
It multiplies the denotations of all tensors together.

On the last line of \cref{fig:named-denote}, the index that \lstinline{denote} passes to \lstinline{denotes} is the \emph{union} (written \lstinline{∪ᵢ}) of \lstinline{idx} (whose shape is \lstinline{zs}) and \lstinline{idx'} (whose shape is \lstinline{⋃ xss ∖ zs}).
Note that these two shapes are disjoint (written \lstinline{⊥}), meaning they share no index variable name.
Hence they are compatible, and that is why it makes sense to take the union of the two shapes and of the two indices.
Since the shape \lstinline{zs ∪ (⋃ xss ∖ zs)} is equivalent to the shape \lstinline{⋃ xss}, we can convert \lstinline{idx ∪ᵢ idx'} from the former shape to the latter.
With the proper indices, \lstinline{denotes} can find values in each input tensor and complete the multiplication.

\subsection{Translating from Unnamed to Named}
\label{section:syntax-translate}

The unnamed tensor language models what tensor libraries implement, but the named tensor language is what we mechanize.
It is thus reasonable to ask, as with all language mechanizations, whether our \lstinline{TExpr} type faithfully captures the expressivity, complexity, and interaction of tensor operations.
To address this question, we translate the unnamed language to the named language.
Our translation takes an unnamed tensor expression~$E$, say of shape $(s_1,\dotsc,s_r)$, and also takes a sequence of distinct names $x_1,\dotsc,x_r$.
It~produces a named tensor expression~\lstinline{t}, of shape $\text{\lstinline{dims}} = \smash[t]{x_1^{(s_1)} \dotsm x_r^{(s_r)}}$, such that for any environment~$\gamma$ and index~$I$ we have
\begin{equation}
\label{e:trans-correct}
    \denot{E}_\gamma [I] = \text{\small\ttfamily denote t env idx}
\end{equation}
where \lstinline{env} consists of the unnamed tensors in~$\gamma$ stored in de Bruijn style,
and \lstinline{idx}${}={}$\lstinline{NIndex.fromUnnamed dims neq }$I$, where
\lstinline{neq} is the proof of \lstinline{Unique dims}.

Consider how the translation should work on the expression $E = \einsum(M_{ij}, M_{kj} \to ik)$ from \cref{section:intro}.
Say that $M$ has shape $(2,3)$, so $E$ has shape $(2,2)$, and the given sequence of distinct names (that is, the domain of \lstinline{idx} in \labelcref{e:trans-correct}) is \lstinline{["x","w"]}.
We will translate $E$ to a \lstinline{TExpr} of the form
\begin{lstlisting}
einsum (cons (var [("x",2), ("y",3)] _ _)
        (cons (var [("w",2), ("y",3)] _ _) nil)) _ _
\end{lstlisting}
in which the \lstinline{var} subexpressions are produced by recursively translating the tensor variable expression $M$ at two different sequences of distinct names, \lstinline{["x","y"]} and \lstinline{["w","y"]}.
Producing this result requires mapping from~$i$ to~\lstinline{"x"}, from~$k$ to~\lstinline{"w"}, as well as from the summed-over $j$ to a fresh name~\lstinline{"y"}.
Thus, the translation operates on \emph{name map}s using the helpers \textit{create\_dict} to create mappings, \textit{freshen} to extend mappings, and \textit{rename} to apply mappings.
In particular,
\begin{align}
    \textit{create\_dict}(ik, \text{\small\ttfamily ["x","w"]}) &= \sigma
\\
    \textit{freshen}(ijk, \sigma) &= \sigma'
\\
    \textit{rename}(ij, \sigma') &= \text{\small\ttfamily ["x","y"]}
\end{align}
where we abbreviate the name maps
\begin{align}
    \sigma
    &= \{i \mapsto \text{\small\ttfamily "x"},
         k \mapsto \text{\small\ttfamily "w"}\}
\\
    \sigma'
    &= \{i \mapsto \text{\small\ttfamily "x"},
         k \mapsto \text{\small\ttfamily "w"},
         j \mapsto \text{\small\ttfamily "y"}\},
\end{align}
\lstinline{"y"} is a fresh name, and we omit the sizes and uniqueness proofs of dimensions.
Fresh names also need to be generated at the top level and for the \lstinline{rhs} of each \lstinline{let} expression, so we introduce one more helper $\textit{gendim}$, which can generate fresh names like
\lstinline{[("x",2),("w",2)]} from the shape $(2,2)$.

\begin{figure}
\centering
\begin{equation*}
  \begin{aligned}
    &\mathit{trans}(E) = (\texttt{dims}, \mathit{comp}_{()}(E, \texttt{dims}))\\
    &\quad\text{where}~\texttt{dims} = \gendim{\shapeu{E}}\\[\bigskipamount]
    &\compu{\alpha}{\texttt{[]}} = \mbox{\lstinline|scalar|}~\alpha
    \\
    &\compu{\smash[t]{0^{(s,\dotsc)}}}{\texttt{dims}} = \mbox{\lstinline|zero|}~\texttt{(fromList dims \_)}\\
    &\compu{n\cdot{E}}{\texttt{dims}} = \mbox{\lstinline|scale|}~n~\compu{E}{\texttt{dims}}
    \\
    &\compu{E_{1}+E_{2}}{\texttt{dims}} =\\
    &\quad\mbox{\lstinline|add|}~\compu{E_{1}}{\texttt{dims}}~\compu{E_{2}}{\texttt{dims}}
    \\
    &\compu{\bind{t}{E'}{E}}{\texttt{dims}} =\\
    &\quad\mbox{\lstinline|let|}~\texttt{d}'~\texttt{\_}~\compu{E'}{\texttt{d}'}~\compu[\Gamma;t,\texttt{d}'\texttt{.map Prod.snd}]{E}{\texttt{dims}}\\
    &\quad\text{where}~\texttt{d}' = \gendim{\shapeu{E'}}\\
    &\compu{t}{\texttt{dims}} = \mbox{\lstinline|var|}~\texttt{dims}~\texttt{\_}~\Gamma(t)\\
    &\compu{\expand(E[\vec{x}] \to \vec{z})}{\texttt{dims}} = \\
    &\quad\mbox{\lstinline|expand|}%~(\texttt{xs} := \texttt{d}_{x})~(\texttt{zs} := \texttt{dims})
    ~\compu{E}{\rename{\vec{x}}{\mathit{dict}}}~\texttt{\_}\\
    &\quad\text{where}~\mathit{dict} = \createdict{\vec{z}, \texttt{dims}}\\
    &\compu{\einsum(E_{1}[\vec{x}_{1}],\dotsc \to \vec{z})}{\texttt{dims}} = \\
    &\quad\mbox{\lstinline|einsum|}%~(\texttt{zs} := \rename{\texttt{dims}}{\texttt{neq}})
      ~\texttt[\compu{E_{1}}{\rename{\vec{x}_1}{\mathit{dict}}}\texttt, \dotsc\texttt]~\texttt{\_}~\texttt{\_}\\
    &\quad\text{where}~\mathit{dict} = \textit{freshen}(\vec{x}_1\cup\dotsb, \createdict{\vec{z}, \texttt{dims}})\\
  \end{aligned}
\end{equation*}
\caption{The translator from the unnamed tensor language to the named tensor language}
\label{fig:compiler}
\end{figure}

\Cref{fig:compiler} defines the translation more formally.
See Appendix~\ref{section:proof-trans-correct} for the definitions of the helper functions \textit{shape}, \textit{gendim}, \textit{create\_dict}, \textit{freshen}, and \textit{rename}.

The main function $\data{trans}{}{}$ takes a tensor expression $E$.
It first generates a list \lstinline{dims} of \lstinline{IVar} corresponding to $E$'s shape.
Then it invokes $\data{comp}{}{}$ to recursively translate~$E$.
The $\expand$ and $\einsum$ cases use name maps to consistently replace the names in an index variable sequence with names in \lstinline{dims} as described above.
We omit the lengthy proofs of \lstinline{Unique} and subset relations between \lstinline{Is}es.

If we let \lstinline{t} in \cref{e:trans-correct} be $\compu[\Gamma]{E}{\texttt{dims}}$, then we can prove the equation by induction on~$E$.
%See Appendix~\ref{section:proof-trans-correct} for the proof sketch.

\section{Representing Shapes and Indices}
\label{section:indices}

In this section, we give the details of named shapes.
We first introduce the representation of named shapes and the operations on them in \cref{section:is}.
Then we introduce the operations on named indices in \cref{section:idx}.
Finally we give the properties of these operations in \cref{section:properties}.
These properties are key to the proofs in \cref{section:opt}.

\subsection{Representing Named Shapes}
\label{section:is}

\subsubsection{The definition of \lstinline{Is}}
As explained in \cref{section:named-syntax}, an~\lstinline{Is} represents a finite map from index variable names to dimension sizes.
It can be encoded in many ways, such as as a function or a data structure \cite{friedmanEssentialsProgrammingLanguages2008}.
A proper encoding of~\lstinline{Is} should satisfy the following two requirements.
First, we need an \lstinline{Is} to have the behavior of a finite map: when we look up any index variable name, there should be only one possible dimension size.
Second, since \lstinline{einsum} needs to enumerate indices corresponding to those dimensions that occur in the input but not in the output, we should be able to enumerate all the dimensions from a given \lstinline{Is}.
Based on the two requirements, we define \lstinline{Is} as a structure with two fields: an association list \lstinline{pairs} of \lstinline{IVar} and a proof \lstinline{noDup} that index variable names in \lstinline{pairs} have no duplicates:
\begin{lstlisting}
structure Is where
  pairs : List IVar
  noDup : (pairs.map Prod.fst).Nodup
\end{lstlisting}
We also define the lookup operation of \lstinline{Is}:
\begin{lstlisting}
def lookupPairs (pairs : List IVar) (x : Id) :
  Option UpperBound :=
  (pairs.find? fun dim => dim.1 == x).map Prod.snd

def lookup (xs : Is) (x : Id) : Option UpperBound :=
  lookupPairs xs.pairs x
\end{lstlisting}
Based on \lstinline{lookup}, we can define relations and operations on~\lstinline{Is}.

\subsubsection{Equivalence, subset, compatibility, and disjointness}
Once we define \lstinline{lookup}, we can define the equivalence, subset, and compatibility of two \lstinline{Is}es:
\begin{lstlisting}
def IsEq (xs ys : Is) : Prop :=
  ∀ i , lookup xs i = lookup ys i
notation xs " ≈ " ys => IsEq xs ys
\end{lstlisting}
\begin{lstlisting}
def IsSubset (xs ys : Is) : Prop :=
  ∀ i , ∀ u , lookup xs i = some u →
    lookup ys i = some u
notation xs " ⊑ " ys => IsSubset xs ys
\end{lstlisting}
\begin{lstlisting}
def IsCompat (xs ys : Is) : Prop :=
  ∀ i, (lookup xs i) ∼ᵢ (lookup ys i)
notation xs " ∼ " ys => IsCompat xs ys
\end{lstlisting}
We define the compatibility of \lstinline{Option UpperBound} as follows.
Two \lstinline{Option} \lstinline{UpperBound}s are compatible if they are the same or one of them is \lstinline{none}:
\begin{lstlisting}
inductive CompatVal : Option UpperBound →
  Option UpperBound → Prop where
  | nn : CompatVal none none
  | vv (v : UpperBound) : CompatVal (some v) (some v)
  | nv (v : UpperBound) : CompatVal none (some v)
  | vn (v : UpperBound) : CompatVal (some v) none
notation v₁ " ∼ᵢ " v₂ => CompatVal v₁ v₂
\end{lstlisting}

Two \lstinline{Is}es are disjoint if no index variable name can be found in both of them:
\begin{lstlisting}
def disjointIs (xs ys : Is) : Prop := (i : Id) →
  (lookup xs i = none) ∨ (lookup ys i = none)
notation xs " ⊥ " ys => disjointIs xs ys
\end{lstlisting}
It is easy to see that \lstinline{xs ∼ ys} if \lstinline{xs ⊥ ys}.

\subsubsection{Union, intersection, and difference}
We can define the union $\cup$, intersection $\cap$, and difference $\setminus$ of two \lstinline{Is}es.
Their type declarations are as follows:
\begin{lstlisting}
def unionIs (xs ys : Is) (h : xs ∼ ys) : Is
def intersectIs (xs ys : Is) (h : xs ∼ ys) : Is
def minusIs (xs ys : Is) : Is
\end{lstlisting}
All three functions can be defined by recursion on \lstinline{xs.pairs}, following the usual definitions of union, intersection, and difference.
For \lstinline{∪} and \lstinline{∩}, two \lstinline{Is}es should be compatible to avoid the cases where one index variable name is associated with two different sizes.
For \lstinline{xs ∖ ys}, we simply remove an \lstinline{IVar} from \lstinline{xs} if its index variable name is found in \lstinline{ys}.

\subsubsection{Union, intersection, and difference of dimension sizes}
We can also define the union, intersection, and difference of two \lstinline{Option UpperBound}s.
Their types follow the same pattern of the operations on \lstinline{Is}: union and intersection require a proof that their arguments are compatible, while difference does not.
The union of \lstinline{v₁} and \lstinline{v₂} is \lstinline{some v} if either is \lstinline{some v}, or \lstinline{none} otherwise.
The intersection is \lstinline{some v} if both are \lstinline{some v}, or \lstinline{none} otherwise.
The difference is \lstinline{none} if \lstinline{v₂} is \lstinline{some u}, or \lstinline{v₁} otherwise.
We give their types and notations here.
These operations make it possible to prove the distributivity of \lstinline{lookup}, as can be seen in \cref{thm:is-op-dist}.
\begin{lstlisting}[belowskip=0pt]
def unionVal (v₁ v₂ : Option UpperBound)
  (h : v₁ ∼ᵢ v₂) : Option UpperBound
notation v₁ " ∪ᵤ " v₂ => unionVal v₁ v₂

def intersectVal (v₁ v₂ : Option UpperBound)
  (h : v₁ ∼ᵢ v₂) : Option UpperBound
notation v₁ " ∩ᵤ " v₂ => intersectVal v₁ v₂

def minusVal (v₁ v₂ : Option UpperBound) :
  Option UpperBound
notation v₁ " ∖ᵤ " v₂ => minusVal v₁ v₂
\end{lstlisting}

\subsection{Representing Named Indices}
\label{section:idx}

\subsubsection{Equivalence of named indices}
As introduced in \cref{section:named-denote}, named indices are represented as a structure with a function field.
Two named indices are equivalent if they are equal pointwise:
\begin{lstlisting}[belowskip=0pt]
def NIndex.Compat {xs : Is} (idx₁ idx₂ : NIndex xs) :
  Prop := ∀ {x : Id} {v : UpperBound}
    (hx : lookup xs x = some v),
    idx₁.get hx = idx₂.get hx
notation idx₁ " ≈ᵢ " idx₂ => NIndex.Compat idx₁ idx₂
\end{lstlisting}

\subsubsection{Conversion between named indices}
Given \lstinline{xs, ys : Is} and \lstinline{idx : NIndex xs}, if \lstinline{xs ≈ ys}, we can convert \lstinline{idx} to an \lstinline{NIndex ys}:
\begin{lstlisting}
def NIndex.eq_cast {xs ys : Is} (idx : NIndex xs)
  (xs_ys : xs ≈ ys) : NIndex ys
notation idx " →ᵢ " xy => NIndex.eq_cast idx xy
\end{lstlisting}
If \lstinline{ys ⊑ xs}, we can also convert \lstinline{idx} to an \lstinline{NIndex ys} by removing all dimensions that occur in \lstinline{xs} but not in \lstinline{ys}:
\begin{lstlisting}[belowskip=0pt]
def NIndex.subst_cast {xs ys : Is} (idx : NIndex xs)
  (ys_xs : ys ⊑ xs) : NIndex ys
notation idx " ↓ᵢ " yx => NIndex.subst_cast idx yx
\end{lstlisting}

\subsubsection{Conversion between named and unnamed indices}
In \cref{section:named-denote}, we already know that \lstinline{UIndex} is a sequence of natural numbers in terms of an unnamed shape \lstinline{dims}.
Given the order of dimensions, we can convert between named and unnamed indices:
\begin{lstlisting}
def NIndex.fromUnnamed
  (xs : List Dim) (neq : Unique xs)
  UIndex (xs.map Prod.snd) :
  NIndex (fromList xs neq)

def NIndex.toUnnamed
  (xs : List Dim) (neq : Unique xs)
  NIndex (fromList xs neq) :
  UIndex (xs.map Prod.snd)
\end{lstlisting}
These two functions can be defined by recursion on \lstinline{xs}.

\subsection{Properties of Named Shapes and Indices}
\label{section:properties}

It is easy to see that named shapes have the following properties by the definitions of $\approx$, $\sqsubseteq$, and $\sim$:
\begin{theorem}\label{thm:equiv-properties}
The equivalence relation $\approx$ of two named shapes is reflexive, symmetric, and transitive.
\end{theorem}
\begin{theorem}\label{thm:subst-properties}
The subset relation $\sqsubseteq$ of two named shapes is reflexive and transitive.
\end{theorem}
\begin{theorem}\label{thm:compat-properties}
The compatibility relation $\sim$ of two named shapes is reflexive and symmetric.
\end{theorem}

We can also prove that looking up in the results of $\cup$, $\cap$, and $\setminus$ can be obtained by combining the results of looking up in the operands of $\cup$, $\cap$, and $\setminus$:
\begin{theorem}\label{thm:is-op-dist}
The operation $\lookup$ distributes over union, intersection, and difference of named shapes.
\begin{align}
  \lookup~(\texttt{xs} \cup \texttt{ys})~x &= \lookup~\texttt{xs}~x \cup_{u} \lookup~\texttt{ys}~x\\
  \lookup~(\texttt{xs} \cap \texttt{ys})~x &= \lookup~\texttt{xs}~x \cap_{u} \lookup~\texttt{ys}~x\\
  \lookup~(\texttt{xs} \setminus \texttt{ys})~x &= \lookup~\texttt{xs}~x \setminus_{u} \lookup~\texttt{ys}~x
\end{align}
\end{theorem}
This theorem can be proved by induction on \lstinline{xs.pairs}.
It~is key to proving \cref{thm:einsum-expand} without explicitly defining permutation of index variable sequences.

An important property of named indices is their equality:
\begin{theorem}\label{thm:index-ext}
  Given a named shape \lstinline{xs} and two indices \lstinline{idx₁}, \lstinline{idx₂} of \lstinline{xs}, $\texttt{idx}_{1} = \texttt{idx}_{2}$ if $\texttt{idx}_{1} \approx_{i} \texttt{idx}_{2}$.
\end{theorem}
To prove this property, we use the functional definitions of \lstinline{idx₁}, \lstinline{idx₂} and function extensionality.
With this property, when we pass a tensor of shape \lstinline{xs} and two equivalent indices of \lstinline{xs} to \lstinline{denote} respectively, we can rewrite one index to the other and thus prove that the two sides are equal to each other:
\begin{lstlisting}[belowskip=0pt]
theorem denote_congr {S : Type} [Semiring S]
  {Γ : Context} {xs : Is}
  (t : TExpr Γ xs) (env : Env (S := S) Γ)
  (idx₁ idx₂ : NIndex xs) (h : idx₁ ≈ᵢ idx₂) :
  denote t env idx₁ = denote t env idx₂
\end{lstlisting}

\section{Optimization by Rewriting}
\label{section:opt}

Having defined the named tensor language and proved the properties of named shapes and indices, we can finally prove some rewriting rules of $\expand$ and $\einsum$ correct.
In \cref{section:expand-fuse,section:einsum-expand}, we present the informal statements of our theorems and sketch their mechanized proofs.
We discuss some details about the mechanized proofs in \cref{section:mechanization}.

Thanks to the fact that all shapes are named and satisfy the necessary equivalence, subset, or compatibility relations, these theorems can be stated and proven mostly using set-theoretical properties of the operations of named shapes.
There is no need to rename or explicitly permute dimensions.

\subsection{The Fusion of $\expand$}
\label{section:expand-fuse}

\begin{theorem}\label{thm:expand-fuse}
Two $\expand$s can be merged into one:
\begin{multline}
\label{e:expand-fuse}
\bigl\llbracket
  \expand\bigl(\expand(E[\vec{x}] \to \vec{x}\vec{y})[\vec{x}\vec{y}] \to \vec{x}\vec{y}\vec{z}\bigr)
  \bigr\rrbracket_{\gamma}[I]
  \\=
\bigl\llbracket
  \expand(E[\vec{x}] \to \vec{x}\vec{y}\vec{z})
  \bigr\rrbracket_{\gamma}[I]
\end{multline}
\end{theorem}
Named tensors help us avoid the bookkeeping of renaming and permutation in the statement and proof of this theorem.
The informal statement \labelcref{e:expand-fuse} contains the operand
\[\expand(E[\vec{x}] \to \vec{x}\vec{y})[\vec{x}\vec{y}],\]
but unnamed tensors would call for the more general
\[\expand(E[\vec{x}] \to \vec{x}\vec{y})[\vec{u}\vec{v}],\]
where $\vec{x}$ and $\vec{y}$ are renamed to $\vec{u}$ and $\vec{v}$ respectively.
To prove the theorem in the unnamed language, we would need to do the bookkeeping of renaming.
In the named version, since names are intrinsic to shapes and we define an equivalence relation between named shapes, we do not need to worry about dimension renaming.

Similarly, the output shape $\vec{x}\vec{y}\vec{z}$ in~\labelcref{e:expand-fuse} is chosen as a representative, but the actual output shape can be $\vec{y}\vec{x}\vec{z}$, $\vec{z}\vec{x}\vec{y}$, or any other permutation of $\vec{x}\vec{y}\vec{z}$.\pagebreak[1]
Since the shape of an unnamed tensor is an ordered list of numbers, we have to encode permutation if we want to prove the theorem about unnamed tensors.
In the named version, since shapes are finite maps from names to dimension sizes, we can avoid permutation.

\begin{proof}[Proof sketch]
  Following the denotational semantics, the equation reduces to the form
  \[\llbracket{E}\rrbracket_{\gamma}[\dots] = \llbracket{E}\rrbracket_{\gamma}[\dots]\]
  For the left hand side, we first convert $I$ from a named index of $\vec{x}\vec{y}\vec{z}$ to $\vec{x}\vec{y}$, then to $\vec{x}$.
  For the right hand side, we convert $I$ directly from $\vec{x}\vec{y}\vec{z}$ to $\vec{x}$.
  Both conversions remove the same dimensions from~$I$, so they leave the same dimensions in $I$.
  Therefore, both sides have the same index.
  Since $E$, $\gamma$, and named indices are the same, the two sides are equal.
\end{proof}

The mechanized theorem statement adds the necessary side conditions.
\begin{lstlisting}
theorem expand_fuse {S : Type} [Semiring S]
  {Γ : Context} {xs ys zs : Is}
  (t : TExpr Γ xs) (xy : xs ⊑ ys) (yz : ys ⊑ zs)
  (env : Env (S := S) Γ) (idx : NIndex zs) :
  denote (.expand (.expand t xy) yz) env idx =
  denote (.expand t (IsSubset_trans xy yz)) env idx
\end{lstlisting}
Since the subset relation between input and output shapes of \lstinline{expand} is guaranteed by the well-formedness of named tensor expressions, we can specify the output shape of \lstinline{expand} directly rather than using the juxtaposition $\vec{x}\vec{y}$.
We do not need to worry about the dimension renaming allowed by the unnamed tensor language when nesting expressions, such as $\expand\bigl(\expand(E[\vec{x}] \to \vec{x}\vec{y})[\vec{x}\vec{y}'] \to \vec{x}\vec{y}'\vec{z}\bigr)$, where $|\vec{y}| = |\vec{y}'|$.
In a sense, the translator in \cref{section:syntax-translate} consistently renames the index variables ahead of time.

\subsection{The Commutativity of $\expand$ and $\einsum$}
\label{section:einsum-expand}

\begin{theorem}\label{thm:einsum-expand}
The operations $\einsum$ and $\expand$ commute up to a factor:
\begin{multline}
\label{e:einsum-expand}
\llbracket
  \einsum\bigl(\expand(E[\vec{x}] \to \vec{x}\vec{y})[\vec{x}\vec{y}], F[\vec{z}] \to \vec{w}\bigr)
  \rrbracket_{\gamma}[I]
  \\=
\llbracket
  \expand\bigl((f \cdot \einsum(E[\vec{x}], F[\vec{z}] \to \vec{u}))[\vec{u}] \to \vec{w}\bigr)
  \rrbracket_{\gamma}[I]
\end{multline}
where
\begin{align}
\vec{u} &= \vec{w} \setminus (\vec{y} \setminus \vec{z}), &
\vec{v} &= (\vec{y} \setminus \vec{z}) \setminus \vec{w}, &
f &= \textstyle\prod_{i \in \vec{v}}|i|.
\end{align}
\end{theorem}
As with \cref{thm:expand-fuse}, names intrinsic to tensors help us state and prove this theorem without renaming and permutation bookkeeping.
When $\einsum$ and $\expand$ meet in the unnamed tensor language, the two occurrences of $\vec{x}\vec{y}$ in~\labelcref{e:einsum-expand} may not match, nor the two occurrences of~$\vec{u}$.
Also, the concatenation $\vec{x}\vec{y}$ may be permuted.
To prove the more general theorem in the unnamed language, we would need to interleave bookkeeping with reasoning about sums and products.
In contrast, our named shapes can be thought of as unordered sets; we
visualize them as such in Appendix~\ref{section:sets}.
\begin{proof}[Proof sketch]
The denotation on the left hand side of \labelcref{e:einsum-expand} is a summation over indices of shape $(\vec{x}\vec{y}\cup\vec{z}) \setminus \vec{w}$.
This shape turns out to be the disjoint union (\cref{section:is}) of two shapes: $\vec{v}$ consists of the dimensions that are not used to index any input tensor, and $(\vec{x}\cup\vec{z}) \setminus \vec{u}$ consists of the dimensions that are used to index some input tensors.
The main argument, then, is to split the summation into two nested summations.
One of them ranges over indices of shape~$\vec{v}$, and reduces to the constant factor~$f$ because $\vec{v}$ is unused.

Specifically, using the conditions $\vec{x} \disjoint \vec{y}$ and $\vec{w} \sqsubseteq \vec{x}\vec{y}\cup\vec{z}$, the two sides of \labelcref{e:einsum-expand} separately reduce to
\begin{multline}
\label{e:einsum-expand-split}
      \sum_{
        \delta'\in\dom{\vec{v}}
      }
      \sum_{
        \delta\in\dom{(\vec{x}\cup\vec{z})\setminus\vec{u}}
      }
\begin{array}[t]{@{}l@{}}
        \llbracket E \rrbracket_\gamma [(\delta'\cup\delta\cup\{\vec{w}\mapsto I\})(\vec{x})] \\[.5\jot]\enspace{}\cdot
        \llbracket F \rrbracket_\gamma [(\delta'\cup\delta\cup\{\vec{w}\mapsto I\})(\vec{z})]
\end{array}
\\=
    \Bigl(\prod_{i \in \vec{v}}|i|\Bigr) \cdot
    \sum_{
        \delta\in\dom{(\vec{x}\cup\vec{z})\setminus\vec{u}}
      }
\begin{array}[t]{@{}l@{}}
    \llbracket E \rrbracket_\gamma [(\delta\cup\{\vec{w}\mapsto I\})(\vec{x})] \\[.5\jot]\enspace{}\cdot
    \llbracket F \rrbracket_\gamma [(\delta\cup\{\vec{w}\mapsto I\})(\vec{z})]
\end{array}
\end{multline}
  Since $\vec{v}$ doesn't contain dimensions from $\vec{x}$ or $\vec{z}$, we know $\vec{v}$ is not used to index the denotations of $E$ and $F$.
  Thus, the named index $\delta'$, whose domain is $\vec{v}$, is not used in the first summation in \labelcref{e:einsum-expand-split}.
  Therefore, that summation reduces to the constant factor $\prod_{i \in \vec{v}}|i|$ on the right hand side.
\end{proof}

The mechanized version of \cref{thm:einsum-expand} follows the informal statement and proof above.
Since it requires a lot of side conditions, we use a helper function \lstinline{EinsumExpandCommEq} to produce the desired equation between two denotations.
\begin{lstlisting}
theorem einsum_expand_comm_two
  {S : Type} [Semiring S]
  {Γ : Context} {xs ys zs ws : Is}
  (t₁ : TExpr Γ xs) (t₂ : TExpr Γ zs)
  (xy : xs ⊥ ys)
  (xyz : unionIs xs ys (disjointIs_compat xy) ∼ zs)
  (w_xys : ws ⊑ ⋃ (IssCompat_two xyz))
  (env : Env (S := S) Γ) (idx : NIndex ws) :
  EinsumExpandCommEq t₁ t₂ xy xyz w_xys env idx
\end{lstlisting}

\begin{corollary}\label{lem:einsum-expand}
Summation-free $\einsum$ and $\expand$ commute:
\begin{multline}
  \llbracket
    \einsum\bigl(\expand(E[\vec{x}] \to \vec{x}\vec{y})[\vec{x}\vec{y}], F[\vec{z}] \to \vec{x}\cup\vec{y}\vec{z}\bigr)
    \rrbracket_{\gamma} \\=
  \llbracket
    \expand\bigl(\einsum(E[\vec{x}], F[\vec{z}] \to \vec{x}\cup\vec{z})[\vec{x}\cup\vec{z}] \to \vec{x}\cup\vec{y}\vec{z}\bigr)
    \rrbracket_{\gamma}
\end{multline}
\end{corollary}

\begin{proof}[Proof sketch]
  Since the \lstinline{einsum} expression is summation-free, $\vec{w} \approx (\vec{x}\vec{y} \cup \vec{z})$.
  Therefore, $\vec{v} = (\vec{y} \setminus \vec{z}) \setminus \vec{w} = \vec{y} \setminus (\vec{z} \cup \vec{w}) = \vec{y} \setminus (\vec{z} \cup \vec{x}\vec{y}) = ()$, and $f = 1$.
  By \cref{thm:einsum-expand}, the two sides are equal.
\end{proof}

\subsection{Notes on the Mechanization}
\label{section:mechanization}

We mechanized the proofs in Lean 4.27.0 and Agda 2.7.0.1\@.
For Lean, we used Mathlib's \lstinline{Semiring} infrastructure.
For Agda, we used $\mathbb{N}$ as the scalar value type.
% TODO: maybe mention LOC?
The Lean proof uses helper theorems provided by Mathlib about summation; we have not proven \cref{thm:einsum-expand} in Agda.
Both versions follow the same design as described above.
We took advantage of Agda's customized syntax and hole system to explore the design of the tensor language.
Then we ported the proof to Lean and generalized it to an arbitrary semiring.
We also used Codex to search for related theorems from Mathlib.

Despite the fact that Codex can give proofs of lemmas, it~often generated dense proofs that did not take advantage of lemmas we have proven.
For Agda, it tended to generate large equational reasoning expressions that used a lot of transitivity and symmetry without extracting intermediate steps.
Occasionally it did generate very concise proofs.
When we split the proof into multiple steps, the proofs it generated became more readable.

\section{Related Work}
\label{section:related}

Both $\einsum$ and tensor description languages have a large number of implementations and applications.
We discuss the implementation, optimization, and formalization of $\einsum$.
We also briefly introduce the work on named tensor notation, and on verifying tensor description languages.

% practical use of einsum
\subsection{Implementation and Optimization of $\einsum$}

Different forms of $\einsum$ have been implemented in various languages and libraries.
Matrix multiplication, arguably the most common case of $\einsum$, is fundamental to scientific computing \cite{UpdatedSetBasic2002} and thus has been implemented in many tensor libraries, including the well-known NumPy \cite{harris2020array}, PyTorch \cite{Ansel_PyTorch_2_Faster_2024}, and TensorFlow \cite{tensorflow2015-whitepaper}.
Because matrix multiplication is so fundamental, these libraries also optimize its performance targeting specific hardware or even specific matrix sizes \cite{jhuraniGEMMInterfaceImplementation2013,kjolstadTensorAlgebraCompiler2017}.
We refer readers to \citet{golubMatrixComputations2013} for a more detailed introduction to this topic.

For the general form of $\einsum$, many languages have their own implementations \cite{blacherEfficientPortableEinstein2023,williamsAhwilliaEinsumjl2026,ahlmann-eltzeEinsumEinsteinSummation2026,niuOvershikiHhlo2026}.
There are also implementations that focus more on the algebraic properties of tensors \cite{bernardyDomainspecificTensorLanguages2025,tooby-smithFormalizationPhysicsIndex2024}.
It is well-known that finding the most efficient way (\emph{contraction path}) of computing $\einsum$ is NP-hard \cite{lamOptimizingClassMultidimensional1997}.
Therefore, many libraries for optimizing the performance of $\einsum$ are implemented \cite{g.a.smithOpt_einsumPythonPackage2018,breuerEinsumTreesAbstraction2025,bayraktarCuQuantumSDKHighPerformance2023,dusellBdusellSemiringeinsum2026,ziogasDeinsumPracticallyOptimal2022a}.
In the context of quantum computing, $\einsum$ is often expressed as tensor network contraction.
There are also a lot of efforts to optimize the performance of tensor network contraction \cite{grayQuimbPythonPackage2018,friedQTorchQuantumTensor2018a}.

% correctness of einsum: semantics, fuzzy tests
\subsection{Formalization of $\einsum$}

Even though $\einsum$ has been used by physicists for over a century and adopted by the machine learning community for over a decade, its formalization is pretty recent work.

\Citet{wenigSyntaxSemanticsEinsum2025} give a clear semantics for $\einsum$.
Their formalization, unlike ours, allows repeated index names for both inputs and outputs.
However, they don't consider the interaction between $\einsum$ and other tensor operations or provide a mechanized proof.

\Citet{DBLP:journals/corr/abs-2603-18372} use $\einsum$ expressions for fuzzing sparse tensor compilers, but they don't consider the formalization of $\einsum$ per se.

\Citet{tooby-smithFormalizationPhysicsIndex2024} formalizes the tensor notation used in physics, where the Lean syntax is used to represent index names.
But their formalization does not consider the form of $\einsum$ based on multi-dimensional arrays.
They also provide no rewriting rules for $\einsum$.

\subsection{Named Tensor Notation}

It is not new to name dimensions of tensors.
For example, the definition of $\einsum$ inherently requires names for tensor dimensions.
In tensor network libraries like Quimb \cite{grayQuimbPythonPackage2018}, tensors can be constructed with dimension names.
As discussed by \citet{chiang2023named}, several tensor libraries have implemented this notation.
\Citet{chiang2023named} formalize the idea with semantics and working examples, but their concern is not mechanization.

The mechanization of \citet{tooby-smithFormalizationPhysicsIndex2024} does support named dimensions.
But tensor contraction is done by pairing dimensions.
Only a pair of dimensions can be contracted at one time.
If we want to express $\einsum$ that sums over multiple dimensions, we have to manually choose the contraction order and write $\einsum$ as a nested sequence of contractions.

% tensor language correctness
\subsection{Verification of Tensor Languages}

As discussed above, based on what ``tensor'' means, the formalization of $\einsum$ can also have different meanings.
Therefore, the formalization of $\einsum$ always follows a formalization of a tensor description language, which has a much richer literature \cite{liuVerifiedTensorprogramOptimization2022,10.1145/3747524,aroraTensorRightAutomatedVerification2025,liuVerifiedCompilerFunctional2024,nezamabadiVerifiedVCGVerified2026,tooby-smithFormalizationPhysicsIndex2024}.
It is possible to translate $\einsum$ into these formalized tensor languages by composing the operations in them.
There are two reasons why we do not formalize $\einsum$ this way.
First, most mechanized tensor languages use unnamed tensor notation, which means we still need to find a way to guarantee the correctness of naming dimensions.
Second, translating $\einsum$ into existing operations makes it less obvious to find the rewriting rules.

\section{Conclusion}
\label{section:conclusion}

We mechanize a tensor language that supports $\einsum$ and $\expand$ with named tensor notation.
We give the syntax and the denotational semantics of the unnamed version of the language on paper and the named version mechanized in a proof assistant.
We provide a translation from the unnamed tensor language to the named tensor language.
We prove that the translation preserves the semantics of the unnamed language.
We discuss the techniques required to support named tensors and how they are used to mechanize $\einsum$.
We prove the rewriting rules about $\einsum$ and $\expand$ that can improve the efficiency of $\einsum$ calculation.

\bibliographystyle{ACM-Reference-Format}
\bibliography{main}

\appendix

%\section{Definitions and Proof about the Translator}
%\subsection{The helper functions of \textit{trans}}
\section{Translation Helper Functions}\label{section:proof-trans-correct}
\subsection{\textit{shape}}

The shape of an unnamed tensor is returned by \textit{shape}, which takes a context $\Gamma$ and an unnamed tensor expression $E$.
It returns a list of upper bounds.
The definition is shown below.
{
% \begin{figure}
% \centering
\begin{equation*}
  \begin{aligned}
    &\shapeu{\alpha} = ()\\
    &\shapeu{0^{(s,\dotsc)}} = (s,\dotsc)\\
    &\shapeu{n\cdot{E}} = \shapeu{E}\\
    &\shapeu{E_{1}+E_{2}} = \shapeu{E_{1}}\\
    &\shapeu{\bind{t}{E_{1}}{E_{2}}} = \shapeu[\Gamma;t,\shapeu{E_{1}}]{E_{2}}\\
    &\shapeu{t} = \Gamma(t)\\
    &\shapeu{\expand(E[\vec{x}] \to \vec{z})} = \lvert\vec{z}\rvert\\
    &\shapeu{\einsum(E_{1}[\vec{x}_{1}],\dotsc,E_{k}[\vec{x}_{k}] \to \vec{z})} = \lvert\vec{z}\rvert\\
  \end{aligned}
\end{equation*}
% \caption{The shape of an unnamed tensor.}
% \label{fig:shape}
% \vspace{-10pt}
% \end{figure}
}

\subsection{\textit{gendim}}

To generate an \lstinline{IVar} list \lstinline{d} and a proof \lstinline{n} that names in \lstinline{d} have no duplicates, we invoke \textit{gendim}.
It takes a list of upper bounds and returns a pair consisting of an \lstinline{IVar} list and a no-duplicate proof.
Here, we rely on a function \textit{gensym} to generate a fresh name and a proof that the name doesn't occur in \lstinline{d}.
\begin{equation*}
  \begin{aligned}
    &\gendim{()} = \texttt{[]}\\
    &\gendim{s:s*} = (x',s):\gendim{s*}\\
    &\quad\text{where}~x'~\text{is a fresh name}
  \end{aligned}
\end{equation*}

\subsection{\textit{create\_dict}}

We represent a name map as an association list.
To construct a name map, \textit{create\_dict} takes an index variable sequence $\vec{x}$ and a corresponding \lstinline{IVar} list \lstinline{d}.
It zips the names of $\vec{x}$ and \lstinline{d} to get a map from names in the former to the latter.
Here, we use $\vec{x}.\mathit{vars}$ to represent the list of names in $\vec{x}$.
The definition of \textit{create\_dict} is shown below.
{
  \begin{equation*}
    \begin{aligned}
      &\createdict{\vec{x}, \texttt{d}} = \mathit{zip}(\vec{x}.\mathit{vars}, \texttt{d.map Prod.fst})
    \end{aligned}
  \end{equation*}
}

\subsection{\textit{freshen}}

The function \textit{freshen} takes an index variable sequence and a name map.
It returns a name map.
\begin{equation*}
  \begin{aligned}
    &\textit{freshen}((), \mathit{dict}) = \mathit{dict}\\
    &\textit{freshen}(\ivar{x}{s}\vec{x}, \mathit{dict}) ~|~x \in \mathit{dict} =
      \textit{freshen}(\vec{x}, \mathit{dict})\\
    &\textit{freshen}(\ivar{x}{s}\vec{x}, \mathit{dict}) ~|~x \not{\in} \mathit{dict} =
      \textit{freshen}(\vec{x}, \mathit{dict}.\mathit{insert}((x, x')))\\
    &\quad\text{where}~x'~\text{is a fresh name}
  \end{aligned}
\end{equation*}

\subsection{\textit{rename}}

The function \textit{rename} takes an index variable sequence $\vec{x}$ and a name map $\mathit{dict}$.
It returns an \lstinline{IVar} list.
Since it is guaranteed that every name in $\vec{x}$ occurs in $\mathit{dict}$, \textit{rename} simply looks up the name and replaces it with the associated name.
\begin{equation*}
  \begin{aligned}
    &\rename{()}{\mathit{dict}} = \texttt{[]}\\
    &\rename{\ivar{x}{s}\vec{x}}{\mathit{dict}} =
      (\mathit{dict}(x),s):\rename{\vec{x}}{\mathit{dict}}
  \end{aligned}
\end{equation*}

\begin{comment}
\subsection{Proof of \cref{thm:trans-correct}}
\begin{proof}[Proof sketch]
  This theorem can be proved by induction on $E$.
  For example, for the $\einsum$ case, the denotational semantics of both sides reduce to a sum product.
  Since the unnamed shapes and named shapes can be converted to each other by \lstinline{NIndex.fromUnnamed} and \lstinline{NIndex.toUnnamed}, the summed dimensions and multiplied dimensions have a one-to-one map.
  By induction hypothesis, for each input tensor of $\einsum$, its denotational semantics before and after \textit{trans} is also equal.
  Therefore, the sum products of the two denotational semantics are equal to each other.
  The other simpler cases can also be proved in a similar way.
\end{proof}
\end{comment}

\section{Visualizing Named Shapes in \cref{thm:einsum-expand}}
\label{section:sets}

The reader may find the picture below useful for tracking named shapes and their relations in \cref{thm:einsum-expand}.

\definecolor{x}{cmyk}{0,.5,1,0}
\definecolor{y}{cmyk}{1,1,0,0}
\definecolor{z}{cmyk}{0,1,0,0}

\hfil
\begin{tikzpicture}
    \node [inner sep=0, anchor=center] {\includegraphics[trim=12 36 12 36]{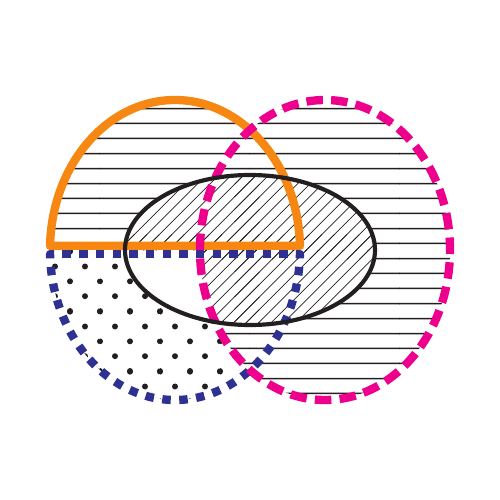}};
    \begin{scope}[every node/.style={fill=white, inner sep=1.5pt}]
        \node at (-3pc,6.2pc) [inner ysep=1pt, text=x] {$\vec{x}$};
        \node at (-3pc,-6.1pc) [inner ysep=1pt, text=y] {$\vec{y}$};
        \node at (8pc,.1pc) [inner xsep=1pt, text=z] {$\vec{z}$};
        \node at (5pc,.1pc) [inner xsep=1pt] {$\vec{w}$};
        \node at (-4.8pc,-3.5pc) {$\vec{v}$};
        \node at (3.2pc,1pc) {$\vec{u}$};
        \node at (3.6pc,3.9pc) {$(\vec{x}\cup\vec{z}) \setminus \vec{u}$};
    \end{scope}
\end{tikzpicture}

\end{document}